%% file: main.tex
\documentclass[sigconf,nonacm=true]{acmart}

\usepackage[utf8]{inputenc}
\usepackage{graphicx}
\usepackage{fancyvrb}
\usepackage{array}
\usepackage{algpseudocode}
\usepackage{cprotect}
\usepackage{makecell}
\usepackage{multirow}
\usepackage{adjustbox}
\usepackage{xcolor,colortbl}
\usepackage{algorithm}
\usepackage{algorithmicx}
\usepackage{tabularx}
\usepackage{makecell}
\usepackage{subcaption}
\usepackage{mathtools}

\newcommand{\annotation}[1]{[[[#1]]]}

    \newcommand{\amit}[1]{\textbf{\small\textcolor{blue}{\annotation{(Amit)~#1}}}{\typeout{#1}}}
    
\newcommand{\eat}[1]{}
\makeatletter
\def\old@comma{,}
\catcode`\,=13
\def,{%
  \ifmmode%
    \old@comma\discretionary{}{}{}%
  \else%
    \old@comma%
  \fi%
}
\makeatother
\theoremstyle{definition}
\newtheorem{observation}[theorem]{Observation}    
\newcommand{\system}{SHARQ}

\newcommand{\df}{D}

\newcommand{\rulesdf}{R}

\newcommand{\tuples}{T}
\newcommand{\srule}{r}
\newcommand{\element}{e}

\newcommand{\column}{a}

\newcommand{\columns}{\mathcal{A}}

\newcommand{\ruleelem}{{\mathcal{E}}({\srule})}
\newcommand{\allelems}{{\mathcal{E}}({\df})}
\newcommand{\elemsshort}{\mathcal{E}}

\newcommand{\ifunc}{I}

\newcommand{\coalition}{S}

\newcommand{\validcoalitions}{\mathcal{C}(e,E)}
\newcommand{\algcoalitions}{\mathcal{C}^*_e}

\newcommand{\optsys}{\system{}^{*}}
\newcommand{\sysopt}{$\system{}^{*}$}
\newcommand{\optsyssingle}{\optsys_{(E, R)}(\element)}
\newcommand{\naivesingle}{\system_{(E, R)}(\element)}

\newcommand{\dfrules}{\rulesdf_\df}

\newcommand{\optcoalitions}{\mathcal{C}^*}

\newcommand{\rulescoresingle}{R\text{-}SHARQ_{(E,R)}(r)}

\newcommand{\attrscoresingle}{A\text{-}\system_{(E,R)}(a)}

\newcommand{\rev}[1]{{\leavevmode\color{black}{#1}}}
\newcommand{\fix}[1]{{\leavevmode\color{black}{#1}}}

\begin{document}

\title{\system{}: Explainability Framework for Association Rules on Relational Data}

\author{Hadar Ben-Efraim}
\affiliation{%
  \institution{Bar-Ilan University}
  %\streetaddress{P.O. Box 1212}
  %\city{Dublin}
  %\state{Ireland}
  %\postcode{43017-6221}
}
\email{hadarib@mail.biu.ac.il}

\author{Susan B. Davidson}
\affiliation{%
  \institution{University of Pennsylvania}
  %\streetaddress{P.O. Box 1212}
  %\city{Dublin}
  %\state{Ireland}
  %\postcode{43017-6221}
}
\email{susan@cis.upenn.edu}

\author{Amit Somech}
\affiliation{%
  \institution{Bar-Ilan University}
  %\streetaddress{P.O. Box 1212}
  %\city{Dublin}
  %\state{Ireland}
  %\postcode{43017-6221}
}
\email{somecha@cs.biu.ac.il}

%%
%% By default, the full list of authors will be used in the page
%% headers. Often, this list is too long, and will overlap
%% other information printed in the page headers. This command allows
%% the author to define a more concise list
%% of authors' names for this purpose.
%\renewcommand{\shortauthors}{Somech and Ben-Efraim, et al.}

\begin{sloppypar}
%%
%% The abstract is a short summary of the work to be presented in the
%% article.

\input{abstract}

%%
%% Keywords. The author(s) should pick words that accurately describe
%% the work being presented. Separate the keywords with commas.

%% A "teaser" image appears between the author and affiliation
%% information and the body of the document, and typically spans the
%% page.
%\hadar{do we need this?}

\maketitle

    %\graphicspath{ {./images/} }

    %%%%%%%%%%%%%%%%%%%%%
    %      INPUTS       %
    %%%%%%%%%%%%%%%%%%%%%
    
%    \input{introduction}
    \input{alt-intro}

    \input{model-shap}

    \input{algorithm}

    \input{applications}

    \input{experiments}

\input{related}

    \input{conclusions}

\clearpage
\bibliographystyle{ACM-Reference-Format}
\bibliography{references}
\clearpage

\end{sloppypar}
\end{document}
\endinput

%% file: abstract.tex
\begin{abstract}
Association rules are an important technique for gaining insights over large relational datasets consisting of tuples of elements (i.e. attribute-value pairs).  However, it is difficult to explain the relative importance of data elements with respect to the rules in which they appear. 
This paper develops a measure of an element's contribution to a set of association rules based on Shapley values, denoted SHARQ (ShApley Rules Quantification). As is the case with many Shapely-based computations, the cost of a naive calculation of the score is exponential in the number of elements. To that end, we present an efficient framework for computing the exact SHARQ value of a single element whose running time is practically linear in the number of rules. Going one step further, we develop an efficient multi-element SHARQ algorithm which amortizes the cost of the single element SHARQ calculation over a set of elements.  
Based on the definition of SHARQ for elements we describe two additional use-cases for association rules explainability:  rule importance and attribute importance.  Extensive experiments over a novel benchmark dataset containing 45 instances of mined rule sets show the effectiveness of our approach.
\end{abstract}

% not written yes

%% file: alt-intro.tex
\section{introduction}

Rule-based pattern mining over large relational datasets 
consisting of tuples of elements (i.e. attribute-value pairs)  
is one of the most popular tools in a data scientist's toolbox~\cite{diaz2022survey,agrawal1993mining},
as it does not requires training data and produces clear data patterns and insights. It has been proven to be highly useful for analyzing data in many application domains, such as E-commerce~\cite{suchacka2017using,dogan2022fuzzy}, biology~\cite{creighton2003mining}, cyber security~\cite{li2021apriori}, and health~\cite{alam2019model,chang2018knowledge}. In particular, association rules mining was used in several research studies on COVID-19 data,
and provided important insights~\cite{katragadda2021association,tandan2021discovering}. 

Since rule mining tools often return thousands of association rules, which can be overwhelming for users, various techniques have been developed to help manage  rules sets.  Solutions for this include ranking rules by different interestingness functions~\cite{bayardo1999mining,brin1997dynamic,zhang2009survey,freitas1998objective} and visualization techniques~\cite{wong1999visualizing,hahsler2017visualizing}
to help users examine and browse the resulting rule set. However, none of these techniques help users understand the relative importance of the \textit{elements} with respect to the rules in which they appear.
Understanding the relative importance of elements not only enhances an understanding of the rules, but can also help reduce the number of rules as demonstrated in the following example.

\begin{table}[t]

\centering
\small
\ttfamily
    \begin{tabular}{|c|c|l|l|c|c|}
    \hline
    \rowcolor{gray!40}
    \textbf{Age} & \textbf{Educ. num} & \textbf{Relationship} & \textbf{Gender} & \textbf{Hrs-per-week} & \textbf{Income} \\
    \hline
    25 & 7 & Own-child & Female & 40 & $\leq 50K$ \\
    28 & 9 & Husband & Male & 50 & $\geq 50K$\\
    29 & 9 & Unmarried & Male & 40 & $\leq 50K$ \\
    44 & 10 & Husband & Male & 40 & $\leq 50K$ \\
    \hline
\end{tabular}

\caption{Adult~\cite{adults_dataset} Dataset Sample}
\label{tab:adults}    
\end{table}

\begin{example}
\label{exa:intro}
A data analyst, Clarice, is examining the \textit{Adults} dataset~\cite{adults_dataset} which provides demographic information on individuals (see Table~\ref{tab:adults} for a small sample of this dataset). 
Clarice is interested in the associations between the elements in the dataset, and uses an association rule mining algorithm on the dataset.  She then focuses on the top-4 rules 
\rev{ranked by the IS score~\cite{tan2000interestingness}, a common interestingness measure for association rules that combines both the frequency with which the rule occurs in the dataset (\textit{support}) as well as strength of the relationship between the left- and right-hand sides (\textit{lift}).
The rules, alongside their support, lift, and combined IS scores are depicted in Table~\ref{tab:rules}}.  These rules include a total of six dataset 
{elements}, $e_1$-$e_6$ (see the two left-most columns of Table~\ref{tab:interestingness}). 

Clarice intuitively notices that element $e_1$, (relationship, unmarried), appears to be less important than the others since if it is omitted from $r_1$ and $r_4$ then the same patterns 
still hold in $r_2$ and $r_3$, with roughly the same IS score. Furthermore, rule $r_3$ has fewer elements than $r_2$, so each of its elements, $e_5$ and $e_6$, would appear to proportionally contribute more than $e_2$-$e_4$. 
Despite their high IS score, rules $r_1$ and $r_4$ therefore appear to be redundant.\qed 
\end{example}

In a sense, measuring an element's contribution to a rule set is an \textit{explainability} problem, yet unlike in Machine Learning (ML) where there are a plethora of techniques to explain the results of an ML model~\cite{shrikumar2017learning,sundararajan2017axiomatic,ribeiro2016should,NIPS2017_7062,ribeiro2018anchors} (see \cite{linardatos2021explainable} for a survey), to our knowledge no such framework exists to explain association rules mined from the data. 
Furthermore, generic measures of an element's contribution to a set of rules, e.g. based on the score of the most interesting rule that contains the element or using causality-based notions such as influence~\cite{pearl2009causality,wu2013scorpion}, do not adequately differentiate between elements.

\begin{example}
\label{exa:interestingness} 
Continuing with the example, Clarice wishes \fix{to} gauge the importance of each of $e_1$-$e_6$ 
to the examined rule set. She employs two intuitive, generic measures to assess the contribution of an element $e$ to the rules set:  
$I_{TOP}(e)$, which returns the IS score of the most interesting rule that contains the element $e$, and $Influence(e)$~\cite{pearl2009causality,wu2013scorpion}, which measures the effect of eliminating the element $e$ from the rules set (see Section~\ref{ssec:defs} for exact definitions).
She is frustrated to find that these measures give \textit{identical} scores to $e_1$-$e_6$ , as shown in columns 3 and 4 in Table~\ref{tab:interestingness}.
\qed 
\end{example}

\begin{table*}[t]
\centering
\small
\ttfamily
    \begin{tabularx}{6in}{|c|X|X|c|c|c|}
    \hline
    \rowcolor{gray!40}
    \textbf{ID} & \textbf{Rule LHS} & \textbf{Rule RHS} & \textbf{Support} & \textbf{Lift}&\textbf{IS score ($\times 10^2$)} \\
    \hline
    $r_1$ & (age, 44-53), (hours-per-week, 40-50), (relationship, unmarried) & (income, >=50K) & 0.2 & 5.25 & 105 \\ \hline
    $r_2$ & (age, 44-53), (hours-per-week, 40-50) & (income, >=50K) & 0.25 & 4.08 & 102 \\ \hline
    $r_3$ & (income, <50K) & (age, 31-44) & 0.43 & 2.44 & 105 \\ \hline
    $r_4$ & (income, <50K) & (age, 31-44), (relationship, unmarried)  & 0.21 & 3.33 & 70 \\
    \hline
\end{tabularx}
\captionof{table}{Example rules from the Adult Dataset with corresponding interestingness scores}
\label{tab:rules}
\end{table*}

\begin{table*}[t]
\centering
%\begin{minipage}{\columnwidth}

\centering
\small
\ttfamily
\centering
    \begin{tabular}{|c|l|c|c|c|}
    \hline
    \rowcolor{gray!40}
    \textbf{ID} & \textbf{Dataset Element} & \textbf{$I_{Top}$} & \textbf{$Influence$} & \textbf{\system{}}  \\
    \hline
    $e_1$ & (relationship, unmarried) & 1.05 & 0 & \textbf{$-0.6$}\\
    $e_2$ & (hours-per-week, 40-50) & 1.05 & 0 & \textbf{$-0.05$}\\
    $e_3$ & (age, 44-53) & 1.05 & 0 & \textbf{$-0.05$}\\
    $e_4$ & (income, >=50) & 1.05 & 0 & \textbf{$-0.05$}\\
    $e_5$ & (age, 31-44) & 1.05 & 0 & \textbf{$4.6$}\\
    $e_6$ & (income, <50) & 1.05 & 0 & \textbf{$4.6$}\\

    \hline
\end{tabular}

\caption{Element contribution scores for elements $e_1$-$e_6$ to the rules set in Table~\ref{tab:rules}}
\label{tab:interestingness}
%\vspace{-5mm}
\end{table*}

Following ideas in the context of explaining ML models~\cite{shap_ml,lundberg2020local} and database query results~\cite{livshits2019shapley,davidson2022shapgraph,bertossi2023shapley}, in this paper we develop a notion of \textit{element contribution} to a set of rules based on the game theoretic notion of Shapley values~\cite{original_shapley}, denoted \textit{SHApley Rules Quantification} (\system{}).

Given a set of rules mined from a dataset and a rule \textit{interestingness} score function~\cite{tan2000interestingness,geng2006interestingness}, \system{} captures the frequency of the element in the dataset as well as the variability in interestingness across rules of different lengths when the element is excluded. In doing so, it provides a finer measure of contribution than the generic measures illustrated earlier. For the example above, \system{} clearly differentiates between the different elements with respect to their contribution to the overall interestingness of the rules set compared to $I_{TOP}$ and $Influence$.  This is depicted in the rightmost column in Table~\ref{tab:interestingness}, where $e_1$ has a significantly lower score than the other elements, and $e_5$ and $e_6$ are higher than $e_2-e_4$, as anticipated in Example~\ref{exa:intro}.

Unfortunately, as is the case with Shapley values in general, calculating the SHARQ score of an element can be extremely expensive. We show that, if done naively, the cost is exponential with respect to the size of the dataset (number of elements). However, by reasoning over a set of input rules we are able to develop an exact, optimized \sysopt{} algorithm that is practically linear with respect to the size of the rule set. This makes exact SHARQ score calculations feasible.  
Going one step more, we show that we can further reduce the cost of calculating individual \sysopt{} scores for a set of elements using a multi-element \sysopt{} algorithm, which amortizes the costs over the set in comparison to calculating the \sysopt{} score of each element separately.

\begin{figure*}[t!]
\vspace{-5mm}
    \centering
    %\begin{subfigure}[b]{0.6\linewidth} 
     %   \centering
        \includegraphics[width=0.9\linewidth]{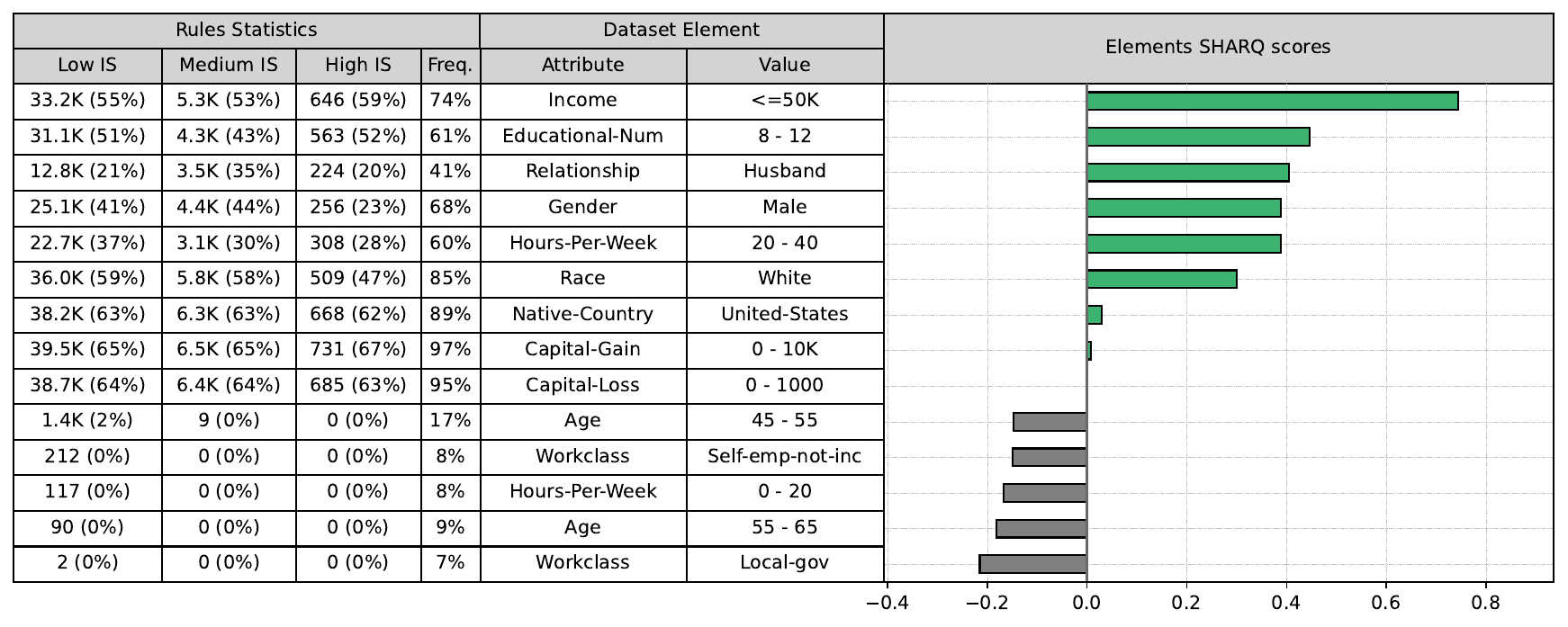}
        \label{fig:usecase1}
    %\end{subfigure}%
    \vspace{-4mm}
    \caption{Example \system{} scores alongside rules statistics, using rules mined from the Adult~\cite{adults_dataset} dataset.}
    \label{fig:basic_ranking}
\end{figure*}

\rev{
Since \system{} scores can be effectively computed for all elements, we explore several explainability use-cases that demonstrate the benefits of our approach. The following example shows a basic use-case of \textit{element importance}, where we analyze the impact of high and low-scoring elements.

\begin{example}
\label{ex:intro_ranking}
Figure~\ref{fig:basic_ranking} shows the \system{} scores for elements contained in an large (85K) rule set mined from the Adult dataset (positive scores are in green, negative scores are in grey). The figure also includes some statistics for each element -- its frequency within the data, and the number of rules it appears in, divided into three interestingness categories (IS): High, Medium, and Low.

We observe that the top five high-scoring elements are frequent and appear mainly in High IS rules. In contrast, the bottom five elements are less frequent and mostly appear in Low IS rules (see Section~\ref{ssec: app_basic} for details).
\end{example}
}

%Once the \system{} scores for the data elements are computed, we further describe several explainability-use cases, showing the benefits of our approach.} The following example illustrates the basic use-case of \textit{element importance} calculation, in which we calculate the \system{} scores for all dataset elements and analyze the impact of both high-scoring and low-scoring elements.

We then show two additional explainability use-cases that are based on element SHARQ scores: rule importance and attribute importance.  
As shown in Example~\ref{exa:intro}, despite their high IS score rules $r_1$  and $r_4$ are redundant of rules $r_2$ and $r_3$, respectively. This could be seen easily since only the top-4 rules were being examined. However, in general, the number of rules above a given interestingness threshold is large, and visually detecting redundant rules becomes infeasible. We show how a notion of rule-level SHARQ scores can be used to narrow the user's focus on a smaller set of important, non-redundant rules.  
Similarly, the SHARQ scores of elements can be used to develop a measure of importance of attributes (features), providing users with a higher-level understanding of which attributes most significantly influence the rule generation process.  

An extensive set of experiments show the effectiveness of our approach. In particular, we show that while a naive calculation of SHARQ is exponential in the size of the dataset and therefore infeasible, the optimized \sysopt{} algorithm grows only linearly in the size of the rules and number of attributes.  
Concretely, computing the SHARQ score for an element using \sysopt{} takes 6.8 seconds on average, compared to hours and even more by the naive SHARQ calculation.
We also show that the  multi-element \sysopt{} algorithm provides an improvement in running time that is proportional to the maximum rule length (12X, in our experiments) over a sequential application of \sysopt{} over a set of elements.
%\susan{Should we phrase this as "We also show that the  multi-element \sysopt{} algorithm provides an improvement in running time that is proportional to the maximum rule length (12X, in our experiments) over a sequential application of \sysopt{} over a set of elements."}
Finally, we show that while generic contribution measures are slightly faster than \sysopt{}, they provide substantially different ranks for the elements, thus cannot be used for approximating \system{}. In contrast, we show that a direct approximation of \sysopt{} based on ~\cite{shap_ml} is significantly better at preserving the element ranking while maintaining similar running times to $I_{TOP}$ and \textit{Influence}.

%Finally, we show that while  approximations of \sysopt{} such as $I_{TOP}$ and \textit{Influence} are faster than \sysopt{}, a direct approximation of \sysopt{}  based on ~\cite{shap_ml} is significantly better at preserving the elements ranking while maintaining similar running times to $I_{TOP}$ and \textit{Influence}.

\paragraph{Contributions}
The contributions of this paper include:
\begin{itemize}
\item
A novel measure of an element's contribution to a set of rules based on Shapley values~\cite{original_shapley} called SHARQ, and theoretical analysis of a naive algorithm based on its definition showing exponential behavior in the size of the dataset.
\item
An efficient algorithm, \sysopt{}, for calculating the \system{} score of a single element, and theoretical analysis showing practically linear behavior in the size of the rule set.   %Although theoretically the size of the rule set  could be exponential in the size of the dataset, in practice the number of rules chosen is much smaller.
\item
An efficient multi-element \sysopt{} for calculating the  \system{} scores for a set of elements, together with an analysis of its complexity. As shown by our use-cases, calculating the SHARQ score for a set of elements arises frequently in practice. 
Processing SHARQ scores jointly, for a set of elements, rather than linearly for each element of the set amortizes the preprocessing cost over all elements and leads to a reduction in running time that is proportional to the maximum rule length
%Processing SHARQ scores for a set of elements rather than linearly for each element of the set amortizes the preprocessing cost over all elements and leads to a 12X reduction in time.
%\susan{Phrase as "Processing SHARQ scores for a set of elements rather than linearly for each element of the set amortizes the preprocessing cost over all elements and leads to a reduction in running time that is proportional to the maximum rule length."}
\item
Use-cases for SHARQ. Building on the notion of element importance, we give a notion of \textit{rule importance} and \textit{attribute importance}.  We also show how our notion of element importance (SHARQ) correlates with the frequency of the element in the dataset, as well as the interestingness of rules in which the element appears. 
\item 
A novel evaluation benchmark containing 45 diverse rule sets mined from four different dataset. Each rules set instance contains a different number of rules, different rule lengths and different number of elements.

\item
Extensive experiments show the effectiveness of our approach: the feasibility of our \sysopt{} algorithm, the effectiveness of our multi-element \sysopt{} algorithm, and the superiority of using a direct approximation of  \sysopt{} rather than generic contribution measures such as $I_{TOP}$ and \textit{Influence}.
\end{itemize}

\paragraph{Outline}
We present
the SHARQ model and problem definition in Section~\ref{sec:model}.  Our single- and multi-element \sysopt{} algorithms are given in Section~\ref{sec: alg}, and use-cases are detailed in Section~\ref{sec:applications}.  Extensive experimental results are presented in Section~\ref{sec: exp}.
After surveying related work in Section~\ref{sec:related}, we close in Section~\ref{sec: concl}.

%% file: model-shap.tex
\section{\system{} Model \& Problem definition}
\label{sec:model}

In this section, we start by defining our model and the notions of rule and rule-set interestingness. We then give a measure of an element's contribution to a rule-set which we call \system{} (Section~\ref{ssec: formula}) and describe its naive implementation. 
%We end by giving an optimized implementation which is still exact (Section~\ref{ssec: optimized}).
\rev{A summary of the notation used throughout the next two sections is given in  Table~\ref{table: notation}. 

\input{notation}}

\subsection{Model and definitions}
\label{ssec:defs}
\eat{We first present definitions for dataset and association rules, and then discuss the notions of rule and rule-set interestingness.}

\paragraph*{Model}
Given a dataset $\df$ with a set of attributes $\columns$ and tuples $\tuples$, a \textit{dataset element} is \fix{an attribute-value pair} $e=(a,v)$, where
$\column \in \columns$ and \fix{value $v  = \Pi_a T$, i.e., the tuple $T$ projected on attribute $a$}.  Given $e= (v,a)$, we will use $attr(e)$ to denote its attribute $a$ and $val(e)$ to denote its value $v$. Correspondingly, $attr(E)$ and $val(E)$ represent the attributes and values (resp.) of a \textit{set} of elements $E$. 
%We denote the value domain of an attribute $a$ by $\valdom$
We denote the set of all elements in the dataset $\df$ by $\allelems$. We will also use $\mathcal{E}(\cdot)$ to denote the elements of a single tuple or of a subset of the data.

We assume that an association rules mining tool~\cite{agrawal1994fast,han2000mining} has been applied to $D$, generating the set of rules $\dfrules$.
A rule $r$ is denoted by 
$E_{LHS} \rightarrow E_{RHS}$,
where $E_{LHS},E_{RHS} \subseteq \allelems$.
Extending $\mathcal{E}(\cdot)$, we use  $\ruleelem= E_{LHS} \cup E_{RHS}$ to denote the set of all elements in the rule $r$.  

An example of rules for the Adult dataset, as described in our running example, is shown in Table~\ref{tab:rules}. 
For example, $r_3$ is the rule (\textit{Income}, \textit{<50K})$\rightarrow$ (\textit{age}, \textit{31-41})
and $\mathcal{E}(r_3)= \; \{(\textit{Income, <50K}) , (\textit{Age, 31-41}) \}$.

\paragraph*{Interestingness of a rule and rules set} 
A rule score function $score(r)$ quantifies the interestingness of a rule $r \in \rulesdf(\df)$.
In our implementation we focus on the IS score~\cite{tan2000interestingness}, which combines two well-known measures: {\em support}~\cite{agrawal1993mining}, which quantifies the frequency of the joint appearance of the rule's elements, and {\em lift}~\cite{brin1997beyond}, which measures the independence deviation of $E_{LHS}$ and $E_{RHS}$: 
$$support(r) = \frac{|\{t | t \in T \wedge \ruleelem \subseteq \mathcal{E}(t)  \}|}{|T|}$$
$$lift(r) = \frac{support(E_{LHS} \cup E_{RHS})}{support(E_{LHS}) \cdot support(E_{RHS})}$$
The full IS score is then defined by: \fix{$$IS(r) \coloneq \sqrt{support(r) \cdot lift(r) }$$} However, our framework can take as input any other measure for rule interestingness as suggested, e.g., in~\cite{geng2006interestingness,freitas1998objective,chandola2007summarization,lin2006outlier,hilderman2013knowledge}.

We will also use a notion for the interestingness of a \textit{set} of rules $R$, defined by
\fix{$$I(R) \coloneq \bigoplus\nolimits_{r \in R} \text{score}(r)$$}
where {\small$\bigoplus$} is an aggregation operation over the individual scores of the rules $r \in R$. In our implementation we use $I(R) \coloneq max_{r \in R}~\text{score}(r)$, but our framework also supports other functions such as summation, average, top-k, etc. See Section~\ref{ssec:ablation} for an empirical comparison. %\amit{complete reference}

\paragraph*{Measuring element contribution to a rules set}
Given a set of dataset  elements $E \subseteq \allelems$ and a set of rules $R \subseteq \rulesdf(\df)$, our goal is to assess the \textit{contribution} of an element $e \in E$ to $I(R)$ the interestingness of $R$. 
There are several possible ways to quantify an element's contribution: For example, one could intuitively define the contribution of element $e$ by the score of the most interesting rule that contains it, which we denote as $I_{TOP}$: 
 $$I_{TOP}(e) \coloneq max_{r \in \{r | r \in R \wedge  e \in r\}} ~\text{score}(r)$$

Another way is to use causality-based notions such as influence~\cite{pearl2009causality,wu2013scorpion}. In our context, {influence} can be defined as the difference in the aggregative interestingness of $R$ when removing the rules that contain $e$. Formally:
$$ Influence(e)\coloneq I(R) - I(R\setminus \{r | r \in R \wedge  e \in r \})$$

However, as demonstrated in Example~\ref{exa:interestingness}, these relatively simple measures do not fully capture the contribution of elements.
We therefore devise a measure of element contribution based on the game theoretic notion of Shapley values~\cite{original_shapley}.
%The Shapley value measures a \textit{player}'s \textit{contribution} to the \textit{utility} of all possible \textit{coalitions}. 
%In our context, the players are elements and the coalitions are rules.
Our measure, SHApley Rules Quantification (\system{}), formally defined in the next subsection,
captures the variability in interestingness across rules of different lengths when the element is excluded. 
Note that, for Example~\ref{exa:interestingness}, \system{} clearly differentiates between the different elements w.r.t. their contribution to the overall interestingness 
as depicted in the rightmost column in Table~\ref{tab:interestingness}, whereas $I_{TOP}$ and $Influence$ do not.
%, where the interestingness measure is implemented as $I(R) \coloneqq max_{r\in R}~IS(r)$).

\subsection{The \system{} Formula}
\label{ssec: formula}

We assume a dataset $\df$ and set of rules $\dfrules$. 
Given a set of elements $E \subseteq \allelems$ and a set of rules $R \subseteq \dfrules$,
our goal is to measure the \textit{contribution} of an element $e \in E$ to the interestingness of $R$. %\footnote{Since $\allelems$ and $\rulesdf(\df)$ may be very large, in our implementation we will focus on subsets of $\allelems$ and $\rulesdf(\df)$, e.g. the most frequent elements or rules within a certain size range.} 
For this, we use the game-theoretic notion of the Shapely value~\cite{original_shapley}, which is widely used in XAI~\cite{vstrumbelj2014explaining,NIPS2017_7062} as well as for data management and exploration tasks~\cite{DavidsonDFKKM22,deutch2020explained,deutch2021explanations} (see Section~\ref{sec:related} for a discussion).

The Shapley value measures a \textit{player}'s \textit{contribution} to the \textit{utility} of all possible \textit{player coalitions}. 
In our adaptation of Shapley values, which we call SHApley Rules Quantification (\system{}),  the ``players'' are elements in $E$ and ``coalitions" are sets of elements with disjoint attributes.  
That is, coalitions model rules %that are constructed with exactly those elements, 
and the contribution of an element is (roughly speaking) the difference in interestingness of rules in which they play a role and those in which they don't play a role.

More precisely, given a set of elements $S \subseteq E$, we use $R_S$ to denote the subset of rules of $R$ that contain \textit{exactly} the elements in $S$: 
$$R_S \coloneq \{\srule|\srule\in R  ~\wedge~\ruleelem = S\}$$
Note that a rule never contains two elements with the same attribute, and therefore $S$ cannot contain two elements with the same attribute.
The \textit{utility} of an element coalition $S$ is then defined as $I(R_S)$, i.e., the \textit{aggregative} interestingness of $R_S$ as defined previously. Naturally, if $R_S= \; \emptyset$ then $I(R_S)= 0$.  

Due to this notion of utility, we consider only \textit{valid} coalitions with respect to an element, i.e. those that can possibly form a rule when the element is included.
Given an element $e$, and a set of elements $E \in \allelems$, we define the  set of \textit{valid} coalitions as follows: 
$$\mathcal{C}(e,E) \coloneq \{S|~S\subseteq E\; \wedge  |attr(\coalition)|=|\coalition|\; \wedge \; attr(e) \notin attr(S)\} $$

%a valid coalition is any 
%subset $S \subseteq E$ that satisfies the condition: $$|attr(\coalition)|=|\coalition|\; \wedge \; attr(e) \notin attr(S)$$
\noindent Namely, all subsets of $E$ that, together with element $e$, do not contain two elements of the same attribute.  
%Note that tuples (and therefore rules) never contain two elements with the same attribute. 

The \system{} score of an element $\element$ in the context of $E$ and $R$ can now be defined as follows:

\begin{small}
$$ SHARQ_{(E,R)}(\element) = \sum_{S\in \validcoalitions} \frac{|S|!(|E|-|S|-1)!}{|E|}\cdot \left(\ifunc\big(R_{S\cup\{\element\}}\big) - \ifunc\big(R_{S}\big)\right)$$
\end{small}

\begin{example}
%Clarice now uses \system{} to calculate the score for elements $e_1$-$e_6$ (shown in the last column of Table~\ref{tab:interestingness}, where the interestingness measure is implemented as $I(R) \coloneqq max_{r\in R}~IS(r)$).
%As expected, she finds that element $e_1$ has a significantly lower score than the other elements, whereas elements $e_5$ and $e_6$ have much higher scores.  Elements $e_2$-$e_4$ have a score which is in-between.
Returning to Clarice's analysis of the \textit{Adults} dataset, the \system{} scores were calculated as follows:  
For element $e_1$ the valid coalitions are any subset of $\{e_2, e_3, e_4, e_5, e_6\}$ that do not contain both $e_4$ and $e_6$ since they both have the same attribute. 
For $\coalition = \{e_2, e_3, e_4\}$ we have 
$\rulesdf_{\coalition} = \{\srule_2\}$ and $\rulesdf_{\coalition\cup\{e_1\}} = \{\srule_1\}$.
As shown in Table~\ref{tab:rules}, $\ifunc(\rulesdf_{\coalition}) = 102$ and $\ifunc\big(\rulesdf_{\coalition\cup\{\element\}}\big) = 105$.
\eat{
The aggregated interestingness of $\rulesdf_{\coalition}$, $\ifunc(\rulesdf_{\coalition})$, is therefore the IS score of $\srule_2$ (see Table~\ref{tab:rules}): $\ifunc(\rulesdf_{\coalition}) = 102$.
 Similarly, $\ifunc\big(\rulesdf_{\coalition\cup\{\element\}}\big) = 105$ (the IS score of $\srule_1$).
 }
Therefore, for $\coalition=\{e_2,e_3, e_4\}$ we get a score of  $\frac{3!(6-3-1)!}{6!}\cdot \left(1.05 - 1.02\right) = 0.05$. We sum the weighted value function difference for the rest of the valid coalitions, and calculate $\system{}(e_1) = -0.6 $. This means that $e_1$ has low contribution (relatively to the other elements) to the rules set formed by $e_1-e_6$ since, when omitted from every valid coalition, there is always another rule with a similar IS score. 

In contrast, $e_5, e_6$ obtain the highest \system{} score of $4.6$. 
For each of these elements, if we add them to a coalition that forms one of the rules $\srule_2, \srule_4$ the interestingness score difference will be negative. For example- $\ifunc(\rulesdf_{\{e_2, e_3, e_4\}\cup\{e_5\}})-\ifunc(\rulesdf_{\{e_1, e_2, e_3\}}) = -102$. For element $e_6$ and coalition $\{e_5\}$ the calculation is $\frac{1!(6-1-1)!}{6!}\cdot\left(105 - 0\right) = 3.5$.
\qed
\end{example}

To find the valid coalitions for an element $e$,
The set $E$ is divided into subsets of elements with the same attribute; elements with attribute $attr(e)$ are omitted.
Coalitions are then formed by choosing either a single element or nothing from each attribute set.

%\paragraph*{Cost Analysis} 
\paragraph*{Number of Coalitions} 
The cost of any application of Shapley values is determined by the number of \textit{coalitions}, which is exponential in the number of \textit{players}. For example, when computing Shapley values in the context of supervised ML explanations~\cite{shap_ml,jethani2021fastshap}, the players are the data \textit{attributes}. When computing Shapley values for database query results~\cite{livshits2019shapley} the players are the dataset \textit{tuples}. In our context, the players are the dataset elements, which is often considerably larger than the number of attributes, and in many datasets can also surpass the number of tuples. 

%The cost of $SHARQ_{(E,R)}(\element)$ is therefore determined by 
The number of valid coalitions for a dataset element $\element$ can be calculated as follows: \fix{Let $E_a$ be the subset of elements in $E$ with attribute $a$, i.e., $attr(e)=a$} 
%Namely, for element $e'=(a',v)$, $e' \in E_a \iff e \in E  \wedge a = a'$.  
Each attribute $a$ has $|E_a| + 1$ options for representation in a coalition - either the attribute does not appear or has one of the values in $\elemsshort_\column$.
Therefore, the number of valid coalitions for $\element$ is $\prod_{a\in \{{attr(E)}\} -\{attr(e)\}} (|E_a| + 1) $.
In the worst case, there are $|T|$ unique elements in each column and $attr(E) = \columns$, hence the cost of calculating the \system{} score for a single element $e$ is $O\left(|T|^{| \columns|}\right)$.

As we show in Section~\ref{ss: opt-naive}, calculating $SHARQ_{(E,R)}(\element)$ becomes infeasible even for small rule sets. We therefore develop an improved formula in which a substantial number of irrelevant element coalitions that do not affect the \system{} score are pruned, and use this to develop efficient algorithms for calculating SHARQ scores.

%\amit{This is a great place to also mention that in ML Shapley, e.g., SHAP, the number of elements is $|\columns|$, and in Benny's Shapley this is $|\tuples|$. Am I correct?  Thus our Shapely calculation is particularly more expensive than SHAP (ML) and Benny's Shapley. }

%% file: notation.tex
\begin{table}[t]
\centering
\small \rev{
\begin{tabularx}{\columnwidth}{|c|X|}
%{|X| c !{\vrule width 1.2pt} c| c|}   
    \hline 
\bf{Variable} & \bf{Meaning}\\ \hline
$\df$ & dataset with attributes $\columns$ and tuples $\tuples$\\
$e= (v,a)$ & dataset element with attribute $a$ and value $v$ \\
$attr(\cdot)$, $val(\cdot)$ & attribute(s), value(s) in tuple/dataset subset\\
$\dfrules$ & set of mined rules of $\df$\\
$\mathcal{E}(\cdot)$ & set of elements in tuple/dataset/ruleset \\
$E \subseteq \mathcal{E}(D)$, $R \subseteq R_D$ & chosen subset of elements, rules \\
$E_\column$ & subset of elements in $E$ with attribute $\column$\\
$r=E_{LHS} \rightarrow E_{RHS}$ & rule \\
$\tau$ & maximum rule size, $|E_{LHS}|+|E_{RHS}|$ \\
$R_S$ & subset of rules containing exactly the elements in $S$\\
$\text{score}(r)$ & score quantifying interestingness of a rule\\
$I(R)$ & interestingness of a set of rules $R \subseteq R_D$\\
$\mathcal{C}(e,E) $ &  set of valid coalitions of $E \subseteq \mathcal{E}(D)$ for element $e$\\
$\mathcal{C^*}(e,E) $ &  optimized set of valid coalitions of $E$ for element $e$\\
$C_r$ & coalitions using elements of rule $r$\\
$SHARQ_{(E,R)}(\element)$ & SHARQ score of $e$ in context
$E \subseteq \mathcal{E}(D)$, $R \subseteq R_D$\\
$SHARQ^*_{(E,R)}(\element)$ & improved formula for $SHARQ_{(E,R)}(\element)$\\
\hline
\end{tabularx}
}
\captionof{table}{\rev{Notation}}
\vspace{-2mm}
\label{table: notation}
\end{table}

%% file: algorithm.tex
%\section{\system{} Optimizations}
\section{Efficient \system{} Algorithms}
\label{sec: alg}
Since calculating the SHARQ score of an element
%$SHARQ_{(E,R)}(\element)$ 
directly is often infeasible, in this section we devise a computational framework, denoted \sysopt{}, for facilitating the \system{} computation while retaining the exact same output as \system{}. First, in Section~\ref{ssec: optimized}, we give an improved formula $SHARQ^*_{(E,R)}(\element)$ in which element coalitions that do not affect the score are pruned. We further prove that the improved formula is equivalent to \system{}, i.e., that $SHARQ^*_{(E,R)}(\element)$ = $SHARQ_{(E,R)}(\element)$. However, in contrast to the original formula in which the number of coalitions is \textit{exponential} in the number of dataset \fix{attributes}, the number of coalitions in the improved formula is several orders of magnitude smaller, bounded by the product of the number of rules and maximum rule size.  This is essentially \textit{linear} in the number of rules since the maximum rule size is typically small (e.g. less than 10)\footnote{This reduction is due to thresholds set on support and interestingness in rule mining algorithms~\cite{han2000mining,agrawal1993mining}.
} 
In Section~\ref{ssec: single} we then show an exact algorithm to calculate $SHARQ^*$ for a single element $e \in E$.

Lastly, since many rule set explanation use cases require the calculation of \system{} scores for \textit{multiple} elements (see Section~\ref{sec:applications}), in Section~\ref{ssec:multi} we describe a multi-element $SHARQ^*$ algorithm. Given a \textit{set} of elements $E$, the algorithm amortizes the cost of materializing the valid element coalitions for \sysopt{} by calculating it \textit{once}, jointly for all elements in $E$.
We show that the multi-element \sysopt algorithm is superior to a sequential application of $SHARQ^*$ over each element of $E$, reduces running time costs by an average of about 12X.  

\subsection{Improved Formula: \sysopt{}}
\label{ssec: optimized}
Our improved formula, \sysopt{}, is based on the observation that an element coalition $S$ has no contribution to the $SHARQ_{(E,R)}(\element)$ calculation
if $R_S = \emptyset$ and $R_S\cup\{\element\}= \; \emptyset$.  
In this case, $\ifunc\big(R_{S\cup\{\element\}}\big) = \ifunc\big(R_{S}\big) = \; 0$. 
Consequently, we can safely restrict the \system{} calculation to coalitions formed from the elements in rules or those with one rule element removed:
\begin{equation}
\label{eq:optcoal1}
\fix{\mathcal{C}^*\coloneq 
  \{S | \exists\srule\in R: (S \subseteq\ruleelem ~\wedge~ |S| \geq|\ruleelem|-1)\}}
\end{equation}

\noindent
Given $e$ and $E$, the \textit{optimized valid} coalitions are thus: 

\begin{equation}
\label{eq:optcoal2}
\mathcal{C}^*(e,E) \coloneq \{S | S \in \mathcal{C}^* \; \wedge S \subseteq E \; \wedge \; attr(e) \notin attr(S) \} 
\end{equation}
The improved formula $SHARQ^*$ then becomes:

\begin{equation}
\label{eq:optimized-sharq}
SHARQ^*_{(E,R)}(\element) = \sum_{S \in \mathcal{C}^*(\element,E)} \frac{|S|!(|E|-|S|-1)!}{|E|}\cdot \left(\ifunc\big(R_{S\cup\{\element\}}\big) - \ifunc\big(R_{S}\big)\right)\end{equation}

\vspace{1mm}
We now show that \system{} and \system{}$^*$ are equivalent.
%\rev{(proof omitted due to space)}. 

\begin{proposition} $\forall e \in E$,  $SHARQ^*_{(E,R)}(e) = SHARQ_{(E,R)}(e)$
\end{proposition}
\eat{More precisely, let
\begin{equation}
\label{eq:s_tag}
\mathcal{C}(\element)\coloneqq 
  \{S| S \in \mathcal{P}(E/attr(e))  ~\wedge~ (\exists r \in R: S \subseteq \ruleelem )\}
\end{equation}
}
\begin{proof}
Given an element $e$, we can divide $\mathcal{C}(\element,E)$ into two sets of coalitions: those that are contained in the elements of some rule, denoted $\mathcal{C}_1(e, E)$;
and those that are not contained in the elements of any rule, denoted $\mathcal{C}_2(e, E)$. We formally define theses subsets as:

$$\begin{array}{ll}
\mathcal{C}_1(e, E) = \{S|&~S\subseteq E\; \wedge  |attr(\coalition)|=|\coalition|\; \wedge \; attr(e) \notin attr(S)\ \wedge\\
& (\exists\srule\in R: S\subseteq\ruleelem)\} 
\\
\mathcal{C}_2(e, E) = \{S|&~S\subseteq E\; \wedge  |attr(\coalition)|=|\coalition|\; \wedge \; attr(e) \notin attr(S)\ \wedge \\ 
& (\forall\srule\in R: S\not\subseteq\ruleelem)\}
\end{array}
$$

We can then rewrite the $SHARQ_{(E, R)}(\element)$ expression:
\begin{equation}
\label{eq:1}
\sum_{S\in \mathcal{C}_1(e, E)} \frac{|S|!(|E|-|S|-1)!}{|E|}\cdot \left(\ifunc\big(R_{S\cup\{\element\}}\big) - \ifunc\big(R_{S}\big)\right)\end{equation}  +
\begin{equation}
\label{eq:2}
\sum_{S\in \mathcal{C}_2(e, E)} \frac{|S|!(|E|-|S|-1)!}{|E|}\cdot \left(\ifunc\big(R_{S\cup\{\element\}}\big) - \ifunc\big(R_{S}\big)\right)
\end{equation}

Observe that for every $S\in \mathcal{C}_2(e, E)$, 
$R_S= R_{(S \cup \{e\})}= \emptyset$ and therefore $I(R_{(S \cup \{e\})})= I(R_S)=0$.
Hence we can simplify $SHARQ_{(E, R)}(\element)$ to Equation ~\ref{eq:1}, which considers only the coalitions in $\mathcal{C}_1(e)$.  

Since attributes are never repeated in rules, $\mathcal{C}_1(e, E)$ can be simplified to: 
$$
\mathcal{C}_1(e, E) = \{S|~(\exists\srule\in R: S\subseteq(\ruleelem \cap E))\; \wedge \; attr(e) \notin attr(S)\} 
$$
%This is because, if $S\subseteq\ruleelem$ for some 
%$\srule\in R$, then $S\subseteq E$ and 
%$|attr(S)|=|S|$ since attributes are never repeated in rules. 
Folding in the definition of $C^*$, $\mathcal{C}^*(e,E)$ can be written as:
$$
\begin{array} {ll}
\mathcal{C}^*(e,E) \coloneq \{S |&(\exists\srule\in R: (S \subseteq (\ruleelem \cap E)) \; \wedge \; attr(e) \notin attr(S) \\
& ~\wedge~ |S| \geq|\ruleelem|-1)
\} 
\end{array}
$$
Therefore $\mathcal{C}^*(e, E) \subseteq \mathcal{C}_1(e, E)$.

Now suppose that $SHARQ^{*}_{(E,R)}(e) \ne SHARQ_{(E,R)}(e)$ for some element $e$. 
%Since $\mathcal{C^*}(e) \subseteq \mathcal{C}_1(e)$, it must be the case that t
Then there is a coalition $S$ in $\mathcal{C}_1(e, E)$
that is not in $\mathcal{C^*}(e, E)$ for which  $\ifunc\big(R_{S\cup\{\element\}}\big) - \ifunc\big(R_{S}\big) \ne 0$.

For this to be true, either $S$ has to form some rule in $R$ but not in $S\cup\{\element\}$, or $S\cup\{\element\}$ has to form a rule in $R$ but not in $S$, or both $S$ and $S\cup\{\element\}$ form rules in $R$. In all cases, $S$ must be in $\mathcal{C^*}(e, E)$ by definition.
Therefore $\forall e, SHARQ^{*}_{(E,R)}(e) = SHARQ_{(E,R)}(e)$.
\end{proof}

%\paragraph*{Cost Analysis}
\paragraph*{Number of Coalitions}
Since coalitions in $\mathcal{C}^*(e,E)$ are formed using rules in $\rulesdf$, each rule $r$ creates either $|\ruleelem| + 1$ coalitions (when $r$ does not contain an element $e'$ with $attr(e)=attr(e')$, \fix{each subset $E \subseteq \ruleelem$ of size $ \geq |\ruleelem|$ is considered }), or the single coalition (when $r$ contains an element $e'$ with $attr(e)=attr(e')$ \fix{the only considered coalition is $\ruleelem - \{e'\}$ }).  %\susan{Since the reviewers didn't complain I don't think we should elaborate further. However, the argument I remember is that by definition, $\mathcal{C}^*$ looks at all subsets of rule elements that remove one element plus the one that contains all elements (if it doesn't have an element with $attr(e)$). And if it does contain an element with $attr(e)$ then only the one coalition with that element removed is considered -- because in all cases you are going to add $e$ back in.  It's just a definitioinal argument.}
Assuming that $\gamma$ percent of the rules do not contain an element with $attr(e)$ and that the maximum rule size is $\tau$,
the size of $\mathcal{C}^*(e,E)$ is at worst
$\left(\gamma \cdot|\rulesdf|\cdot \left(\tau +1\right) + \left((1-\gamma) \cdot|\rulesdf|\right) \right)$, which is
$O\left(|\rulesdf|\cdot \gamma \cdot \tau \right)$. 

In the worst case, in which $\gamma =1$ and $\tau=|\columns|$, the size of $\mathcal{C}^*(e,E)$ is $O \left(|\rulesdf|\cdot |\columns| \right)$.
This is a significant improvement over the the number of coalitions in the non-optimized \system{} score, $O\left(|T|^{| \columns|}\right)$, which is \textit{exponential} w.r.t. the number of rows in the dataset. 
%at most $|\rulesdf|(|\columns| + 1)$. 
%Considering these two factors, the number  $\optsys$ is $O\left(|\rulesdf|\cdot|\columns|\right)$. 
%See that the number of coalitions considered in \sysopt is \textit{linear} w.r.t. the rules set size, whereas the cost of the non-optimized \system{} score, $O\left(|T|^{| \columns|}\right)$, is \textit{exponential} w.r.t. the number of rows in the dataset $T$. 

Note, however, that $|\rulesdf|$ may be larger than $|T|$ for some datasets and rule mining settings (in the worst case, exponential in $|T|$). In our experimental evaluation, %as detailed in Section~\ref{ss: setup}, 
we examined \rev{66} rule sets mined from four different datasets. We found that $|R|$ was indeed larger than $|T|$ in 11 cases, but with only a maximum ratio of $|R|\approx 64\cdot |T|$. % (when using the Isolet~\cite{isolet} dataset). 
Even in this extreme case, the number of coalitions used by the optimized $\optsys$ was 9.5K, compared to 16.7M used by the naive \system{} computation.
%which took hours to compute (we stopped the execution after 8 hours). In comparison, with $\optsys$ we calculate the scores for \text{all} dataset elements in 81 seconds. 

\vspace{2mm}

This optimization allows us to compute \system{} scores in a reasonable amount of time -- \rev{6.6} seconds on average, per element, compared to several hours (or more) by the naive \system{} computation. 
We next present our full \sysopt\ algorithm, which improves the calculation by considering two additional aspects: (1) the retrieval of the optimized valid coalition set $\mathcal{C}^*(e,E)$, and (2) the calculation of the utility function difference $\ifunc\big(R_{S\cup\{\element\}}\big) - \ifunc\big(R_{S}\big)$.

\subsection{The \sysopt{} Algorithm (Single Element)}
\label{ssec: single}
The single element $\optsys$ algorithm, shown in Algorithm~\ref{algo:single}, takes as input \fix{a rule set $R \in R(D)$, a set of elements $E \in \allelems$, and a single element $e \in E$. The algorithm has two parts:}
(1) generating the optimized set of coalitions for $e$, $\mathcal{C}^*(e,E)$, and creating a coalition rules index to reduce the cost of calculating the score (lines 3-13), and (2) incrementally calculating the \sysopt score (lines 14-18 of the CalcSHARQ function).
%\susan{Check the sentence above, it was garbled.}

In the first part, a single pass is done over the rule set \fix{$R$}.
For each rule $r$, the coalitions using elements of that rule, $C_r$, are generated (line 6). To reduce the cost of calculating $\ifunc\big(R_{S\cup\{\element\}}\big) - \ifunc\big(R_{S}\big)$, we incrementally create a \textit{coalition rules index}, $\mathcal{R}$, by adding $r$ to $\mathcal{R}_{S}$ for each coalition $S$ in $C_r$, as well as $\mathcal{R}_{S\cup \{e\}}$ if there is not already an element of type $attr(e)$ in $S$  (lines 4-13). 

In the second part, we compute $\optsyssingle$ using the CalcSHARQ function. This is done by incrementally adding the contribution of $e$ to each coalition $S \in \mathcal{C}^*(e,E)$: %(Line~\ref{ln:shap1}): 
We fetch $\mathcal{R}_S$ and $\mathcal{R}_{S \cup \{e\}}$ from the coalition rules index $\mathcal{R}$ (line~\ref{ln:getrules}), 
then calculate the interestingness difference $\ifunc\big(R_{S\cup\{\element\}}\big) - \ifunc\big(R_{S}\big)$ and multiply by the Shapley factorial coefficient $\frac{|S|!(|E|-|S|-1)!}{|E|}$ (Line~\ref{ln:update-scores}). The sum of these individual contribution scores form the final $\optsyssingle$ score, as indicated in Equation~\ref{eq:optimized-sharq}.

\paragraph{Cost Analysis}
In the first part of the algorithm, 
we iterate through each $r \in \rulesdf$ and form coalitions.  Forming a coalition takes time proportional to the size of $r$, which we assume is $\tau$.  Since there are at most $\tau+1$ coalitions per rule,  
%Since $|\ruleelem| \leq |\columns|$ and assuming that the maximum rule size is $\tau$, 
%generating $\mathcal{C}^*(e,E)$ is in  
forming the coalitions for a single rule (line 4) is $O\left(\tau^2 \right)$, yielding an overall cost of $O\left(|R|\cdot\tau^2 \right)$ for the first part.

%In the worst case, $\tau=|\columns|$ and the cost per rule becomes $O\left(|\columns|^2 \right)$, yielding an overall cost of $O\left(|R|\cdot|\columns|^2 \right)$ for the first part.

In the second part, % (Lines~\ref{ln:sfirst2}-\ref{ln:slast2}), we compute the $\optsyssingle$ score by 
we iterate over $\mathcal{C}^*(e,E)$ to compute the $\optsyssingle$ score.  Assuming that accessing the index $\mathcal{R}$ and calculating the utility scores $I(\cdot)$ are both $O(1)$, the cost of this part is the size of $\mathcal{C}^*(e,E)$. 
The overall cost of the two parts is thus $O\left(|R|\cdot\tau\cdot(\tau +\gamma) \right)$. 
Since $\gamma \leq 1$, this is essentially $O\left(|R|\cdot \tau ^2 \right)$.  
Since rules are typically small ($\tau << |\mathcal{A}|$), we call this ``practically linear" in the number of rules.
However, in the worst case, where $\tau=|\mathcal{A}|$
this becomes $O\left(|R|\cdot |\mathcal{A}| ^2 \right)$.
\eat{Since $\gamma \leq 1$ and $\tau \leq |\mathcal{A}|$, this is bounded by    $O\left(|R|\cdot|\columns|^2 \right)$.}
%The overall cost of the algorithm is therefore dominated by the first part, with a worst case cost of $O\left(|R|\cdot|\columns|^2 \right)$.

\eat{As discussed earlier, this is 
$O\left(|\rulesdf|\cdot r_e \cdot \tau \right)$, which in the worst case is
$O\left(|\rulesdf|\cdot|\columns|\right)$.
The overall cost of both parts is therefore 
$O\left(|\rulesdf|\cdot \tau^2 + |\rulesdf|\cdot r_e \cdot \tau\right)$, or in the worst case
$O\left((|\rulesdf|\cdot|\columns|^2+|\rulesdf|\cdot |\columns|\right)$. }
%(recall the discussion in Section~\ref{ssec: optimized}). 
%\amit{should we use the new expression - $O \left(r_e \cdot|\rulesdf|\cdot \left(\tau +1\right) + \left((1-r_e) \cdot|\rulesdf|\right) \right)$?}

%\amit{ If so the expression should be : $$O\left((\tau+1) \cdot |\rulesdf| +  \left(r_e \cdot|\rulesdf|\cdot \left(\tau +1\right) + \left((1-r_e) \cdot|\rulesdf|\right) \right) \right) $$, which can be simplified to $$ O\left(|\rulesdf|\cdot (r_e\cdot \tau + \tau + 2 )\right)  $$

\vspace{2mm}
Next, we extend our discussion of calculating \system{} scores to a \textit{subset} of elements rather than a single one. We present an effective algorithm for this multi-element case which
significantly reduces the cost of generating the coalitions, $\optcoalitions$, as well as the coalition rules index, $\mathcal{R}$.

\begin{algorithm}[t]
\caption{\textbf{Single Element \sysopt}}
\begin{algorithmic}[1]
    \Statex \textbf{Input: } Rule set $R \subseteq R_D$, element set $E \subseteq \allelems$, single element~$\element \in E$
    \Statex \textbf{Output: } $\optsyssingle$, the SHARQ score of element $\element$
    \State $\algcoalitions \leftarrow \text{ Initialize empty optimized coalitions set } \mathcal{C}^*(e,E) $ \label{ln2:sfirst1}
    \State $\mathcal{R} ~~\leftarrow \text{ Initialize empty coalition-to-rules index}$ \label{ln2:init-dict}
    \For {$\text{rule }\srule \in \left\{r| r \in  R\wedge\ruleelem \subseteq E \right\}$} \label{ln:rulescan1}
        \State $C_r = \{\coalition | \coalition \subseteq \elemsshort(\srule) \wedge |\coalition| \ge |\ruleelem| -1\}$
        \For {$\text{coalition } \coalition \in C_r$}
            \State $\mathcal{R}[S] \leftarrow \mathcal{R}[S] + (\srule)$ \label{ln:update-dict1}
            \If {$attr(e) \notin S$} \label{ln:update-ce-first}
                \State $\algcoalitions \leftarrow \algcoalitions \cup \{S\}$
            \EndIf \label{ln:update-ce-last}
        \EndFor
    \EndFor \label{ln:slast1}
    \State \textbf{return} {CalcSHARQ($e$,~$E$,~$\mathcal{R}$,~$\mathcal{C}^*_e$)}

\Statex   \Function{CalcSHARQ($e$,~$E$,~$\mathcal{R}$,~$\mathcal{C}^*_e$)}{}
\State $\optsyssingle \leftarrow 0$ \label{ln:sfirst2}
        \For {$\text{coalition } \coalition \in \algcoalitions$} \label{ln:shap1}
            \State $R_S,R_{S\cup\{\element\}} \leftarrow \mathcal{R}[S], \mathcal{R}[S\cup\{\element\}]$ \label{ln:getrules}
            \State $\optsyssingle \mathrel{+}=$  
            {\small $\frac{|S|!(|E|-|S|-1)!}{|E|}\cdot \left(\ifunc\big(R_{S\cup\{\element\}}\big) - \ifunc\big(R_{S}\big)\right)$} \label{ln:update-scores}
        \EndFor \label{ln:slast2}
        \State \textbf{return} $\optsyssingle$
    \EndFunction
    
\end{algorithmic}
\label{algo:sharq 1 element}
\label{algo:single}
\end{algorithm}

\subsection{Multi-Element \sysopt{}  Algorithm}
\label{ssec: alg}
\label{ssec:multi}

Many of the use cases of \system{} focus on the scores of \textit{all} dataset elements $\allelems$ (or a large subset thereof) rather than the score of a single element (see Section~\ref{sec:applications}). For example, users might want to reduce the number of elements by dropping those with near zero \system{} contribution over a large rule set.  This in turn will reduce the number of rules, allowing users to more quickly observe interesting patterns.  

Given a set of $E$ elements, one can naively run the single element \sysopt{} algorithm, as described in Algorithm~\ref{algo:sharq 1 element}, and compute $\optsyssingle$ for each $e \in E$. However, this is inefficient, since the pre-processing phase, in which we generate the coalition set $\mathcal{C}^*(e,E)$ and the coalitions-to-rules index $\mathcal{R}$, is repeated $|E|$ times.  In doing so, the same coalition set may be recalculated several times:

\begin{observation} If two elements $e,e'\in E$ have the same attribute, $attr(e)=attr(e')$, then $\mathcal{C}^*(e,E) = \mathcal{C}^*(e',E)  $
\label{obs:multi}
\end{observation}

This observation stems from the definition of $\mathcal{C}^*(e,E)$ (see Equation~\ref{eq:optcoal2}), which excludes from $\optcoalitions$ all coalitions that contain an element with the same attribute as $e$.
We can therefore calculate and save the set of coalitions for each \textit{attribute} $a \in attr(E)$ that appears in some element in $E$ rather than saving it for each \textit{element} $e \in E$, thus achieving significant savings in space and time.
%\susan{Modified slightly: rather than for each \textit{attribute} it is each attribute that appears in some element in $E$, as reflected in line 12 of the Multi-element algorithm.} 
%\amit{Modified slightly w.r.t Susan's comment (also Line 7 in Algo2)}

\vspace{1mm}
The multi-element \sysopt algorithm is shown in Algorithm~\ref{algo:multi}.
The algorithm takes a rule set $\rulesdf$ and a set of elements $E$, and calculates all \system{} scores, for each $e \in E$. 
We first initialize a coalitions index $\mathcal{C}$, which stores the set of coalitions for each attribute (based on Observation~\ref{obs:multi}) and the coalitions-to-rules index $\mathcal{R}$ (Lines~\ref{ln2:preprocessing_Start}-\ref{ln2:init-dict}). 
We then scan the rules and update $\mathcal{R}$ as in Algorithm~\ref{algo:sharq 1 element} as we extract coalitions from each rule (Lines~\ref{ln2:rulescan1}-\ref{ln2:update-dict1}). Next, we update the optimized-coalitions index $\mathcal{C}$ when processing each extracted coalition $S$, adding it to the coalitions set of all relevant attributes, i.e., that are in $attr(E) \setminus attr(S)$ (lines~\ref{ln2:coalupdate_begin}-\ref{ln2:coalupdate_end}). 
Finally, to calculate the \system{} scores for each element $e$ we
first fetch its coalitions set from the coalitions index $\mathcal{C}$ (Line~\ref{ln2:get_coalitions}), then call the CalcSHARQ function from Algorithm~\ref{algo:sharq 1 element} (line~\ref{ln2:slast2}).

\begin{algorithm}[t]
\caption{\textbf{Multi-element \sysopt }  } \label{alg0}
	\begin{algorithmic}[1]
	    \Statex \textbf{Input: } Rule set $R \subseteq R_D$, element set $E \subseteq \allelems$
	    \Statex $\text{\textbf{Output: }} \text{ All SHARQ Scores }\optsyssingle, \forall e \in E$
	\State $\mathcal{C} \leftarrow \text{initialize empty optimized-coalitions \underline{index}}$ \label{ln2:preprocessing_Start}
 \State $\mathcal{R} \leftarrow \text{initialize empty coalition-to-rules index}$ \label{ln2:init-dict}
 \For {$\text{rule }\srule \in \left\{r| r \in  R\wedge\ruleelem \subseteq E \right\}$} \label{ln2:rulescan1}
            \State $C_r = \{\coalition | \coalition \subseteq \elemsshort(\srule) \wedge |\coalition| \ge |\ruleelem| -1\}$
	        \For {$\text{coalition } \coalition \in C_r$} \label{ln2:coalupdate_begin}
    	        \State $\mathcal{R}[S] \leftarrow \mathcal{R}[S] + (\srule)$ \label{ln2:update-dict1}
     
                   \For {$\text{attribute } a \in attr(E) \setminus attr(S)$}
                            \State $\mathcal{C}[a] \leftarrow \mathcal{C}[a] \cup \{S\}$ 
                        
	          \EndFor 
            \EndFor \label{ln2:coalupdate_end}
        \EndFor \label{ln2:preprocessing_end}
	    \For {$\text{element } \element \in E$}
            \State $\mathcal{C}^*_e \leftarrow \mathcal{C}[attr(e)]$ \label{ln2:get_coalitions}
            \State  \textbf{yield} CalcSHARQ($e$,~$E$,~$\mathcal{R}$,~$\mathcal{C}^*_e$) \label{ln2:slast2}
        \EndFor 
	    
	\end{algorithmic}
\label{algo:multi}
\end{algorithm}

\begin{comment}
\begin{figure*}[t!]
\vspace{-5mm}
    \centering
    %\begin{subfigure}[b]{0.6\linewidth} 
     %   \centering
        \includegraphics[width=0.9\linewidth ]{plots/Flows/top_bottom_stats_table_12_7.pdf}
        \label{fig:usecase1}
    %\end{subfigure}%
    \vspace{-4mm}
    \caption{Example \system{} scores alongside rules statistics, using rules mined from the Adult~\cite{adults_dataset} dataset.}
    \label{fig:basic_ranking}
\end{figure*}
\end{comment}

\paragraph*{Cost Analysis.}
In the first part of the algorithm 
we iterate through each $r \in \rulesdf$ and form coalitions in order to incrementally construct the coalitions index and coalitions-to-rules index. 
As mentioned previously, this takes $O\left(\tau^2\right)$ per rule.
In addition to adding $r$ to the coalition-to-rules index, we add each coalition $S$ formed from $r$ to the optimized-coalitions index of each attribute $a$ that appears in $E$ but does not appear in $S$.  
%\susan{The complexity of taking the difference of $attr(E)$ and $attr(S)$ is $|attr(E)|$ using hashing, and $|attr(E|)\cdot |attr(S)$ using unordered sets. }
Assuming that the cost of calculating $attr(E) \setminus attr(S)$ is
$|attr(E)|$, the cost of the first part is
$O\left(|\rulesdf|\cdot (\tau^2  + |attr(E)|)\right)$.
%, which in the worst case is $O\left(|\rulesdf|\cdot (|\columns|^2  + |attr(E)|)\right)$.  

In the second part, after building the index structures, we calculate the \system{} scores by iterating over each element $e \in E$, retrieving the optimized valid coalitions for $e$, and computing the final score. Assuming that accessing the indexes $\mathcal{C}$, $\mathcal{R}$ and calculating the utility scores $I(\cdot)$ are all $O(1)$, the cost of this part is $O\left(|E|\cdot|\rulesdf|\cdot\gamma\tau\right)$. 
Since $\gamma < 1$, this could be simplified to $O\left(|E|\cdot|\rulesdf|\cdot\tau\right)$. 
The total cost of the algorithm is 
therefore $O\left(|\rulesdf|\cdot (\tau^2 + |attr(E)| + |E| \cdot \tau ) \right)$.
Recall that the cost of running the single element algorithm for all elements in $E$ is bounded by  $O\left(|E| \cdot |\rulesdf|\cdot \tau(\tau+1)\right)$.
\fix{Since $|attr(E)| \leq |E|$,  the ratio of the costs of the sequential approach and Algorithm~\ref{algo:multi} is \textit{at least} $\frac
{|E| \cdot  \tau(\tau+1)}{|E| \cdot (\tau +1) + \tau^2}$.
Based on this, we can show that, roughly, Algorithm~\ref{algo:multi} reduces running times by a factor greater than $\frac{|E|\cdot\tau}{|E|+\tau +1}$.} 
%This means that, roughly, Algorithm~\ref{algo:multi} reduces running times by a factor greater $\tau$.

In our experiments (See Section~\ref{ss: all-elements}), we show that the multi-element \sysopt{} algorithm achieves a minimal improvement of \rev{6.7X} and a maximal of \rev{25X}, with an average factor of \rev{13.8X}. %(The value of $\tau$ ranges between 7 and 8 in our experiments' rule sets).  

\eat{Using our SHARQ* algorithm, computing the SHARQ score for an element takes \rev{6.6} seconds on average, compared to hours and even more by the naive SHARQ calculation which considers all valid coalitions.}

%% file: applications.tex
\section{Rules Explainability Use Cases}
\label{sec:applications}

We next illustrate three use cases for association rules explainability using \system{}. As in our running example, we use the Adult~\cite{adults_dataset} dataset; unlike the example, we use the \textit{full} set of rules mined from the data.

Our first use case is \textit{element importance}, described in Section~\ref{ssec: app_basic}. In this use case, we calculate the \system{} scores for all dataset elements and analyze the impact of both high-scoring and low-scoring elements. Building on this analysis, we introduce two additional use cases: \textit{rule importance} (Section~\ref{ssec: app_rule_importance}), where we identify \textit{redundant} rules that, despite their high interestingness scores, are similar to other, \textit{shorter} rules with equivalent scores. Removing these redundant rules helps to narrow the user's focus to a more concise set of significant rules. Finally, the \textit{attribute importance} use case (Section~\ref{ssec:app_feature_importance}) leverages the \system{} scores of elements to determine the significance of attributes relative to the mined rules, thus providing users with a higher-level understanding of which attributes most significantly influence the rule generation process.

\subsection{\system{} Element Importance}
\label{ssec: app_basic}

We apply the Apriori~\cite{agrawal1994fast} rules mining algorithm on the Adult~\cite{adults_dataset}, after discretizing the data by binning numeric columns\footnote{Discretization of numeric data is required by most rules-mining algorithms, which traditionally work on transactional data with discrete items.~\cite{agrawal1993mining,han2000mining,srikant1997mining}}, setting the minimum support threshold to 0.05. This results in a set of 84479 rules, spanning 34 unique dataset elements (see our code repository in~\cite{our_github_repository} for full settings and output).
Next, we calculate the \system{} scores for all individual elements in $E=\mathcal{E}(R)$, i.e., all elements that participate in at least one rule. 

\rev{As briefly introduced in Example~\ref{ex:intro_ranking},}
Figure~\ref{fig:basic_ranking} illustrates the \system{} scores, ordered from high to low. For the analysis of the scores, we provide additional statistics for each dataset element: the frequency of the element, denoted as $freq(e) = \frac{|\text{Rows containing element } e|}{|\text{Total rows}|}\times 100$; and the number of rules in which the element appears, divided into three categories of interestingness (IS): Low IS (0.22 - 0.41), Medium IS (0.41 - 0.59), and High IS (0.59 - 0.78). We also specify the proportion of rules (in parentheses) that fall into each IS category. Out of the total 84,479 rules, 69,776 are classified as \textit{low interestingness}, 13,358 as \textit{medium interestingness}, and 1,345 as \textit{high interestingness}.

%\amit{specify IS score ranges, with a maximum of one digit after the decimal}. 

Analyzing the \system{} scores, we observe that the top five high-scoring elements are highly frequent in the data, each with a frequency exceeding 41\%. Furthermore, these elements prominently appear in High IS rules (more than 20\%) and Medium IS rules (over 35\%). In contrast, the bottom five elements are less common, appearing in less than 17\% of the rows, and are predominantly found in Low IS rules.

Additionally, we note that some highly-frequent elements in the data obtain nearly zero \system{} scores. Examples are \textit{(native-country, United States)}, \textit{(Capital-Gain, 0-10K)}, and \textit{(Capital-Loss, 0-1K)}, with respective frequency values of 89\%, 97\%, and 95\% and \system{} scores of 0.03, 0.007, and 1.3e-5, respectively. This occurs because \system{} scores take into account the importance of an element to a rule, as discussed in Example~\ref{exa:intro}. Despite their high frequency in the data, these elements are considered \textit{non-influential} because removing them leads to shorter, yet equivalently interesting rules.

We next demonstrate how \system{} can be used to identify rule redundancy, thereby allowing the user to focus on a smaller, more concise set of rules.

\begin{figure}[t!]
    \centering
        \includegraphics[width=1\linewidth]{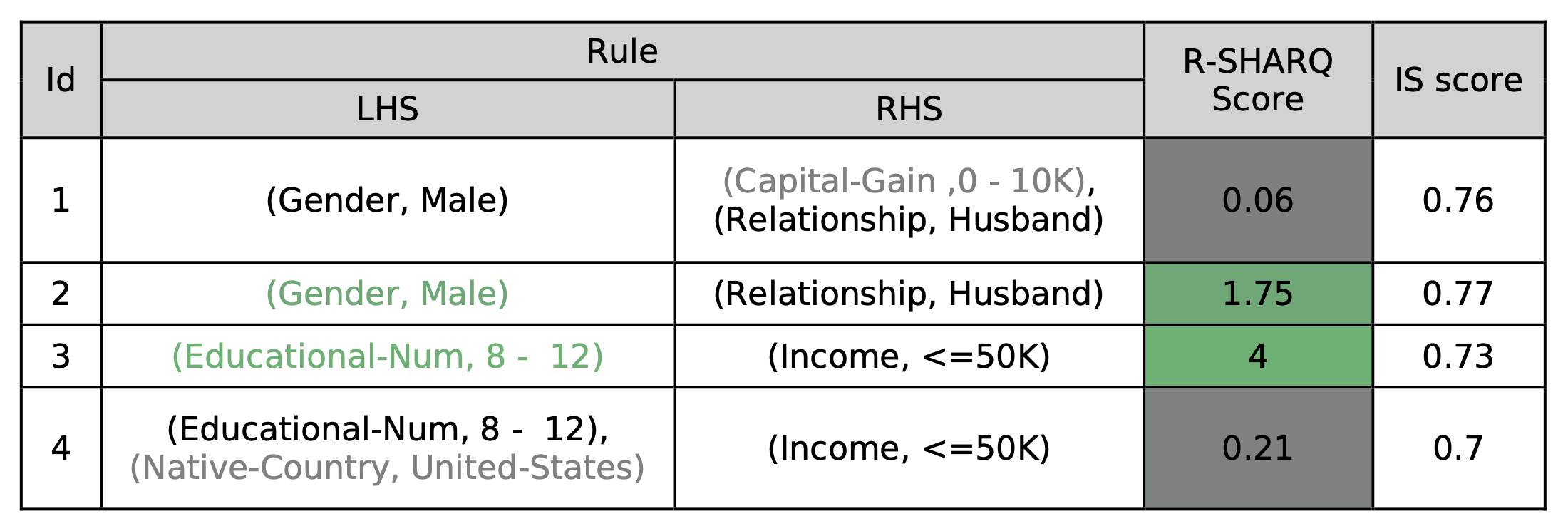}
        \vspace{-6mm}
    \caption{R-SHARQ rule score examples. }
    \label{fig:alt_rules}
\end{figure}

\subsection{Rule Importance: Detecting Redundancy}
\label{ssec: app_rule_importance}

\begin{figure*}[t!]
    \centering
    \vspace{-7mm}
        \includegraphics[width=0.9\linewidth]{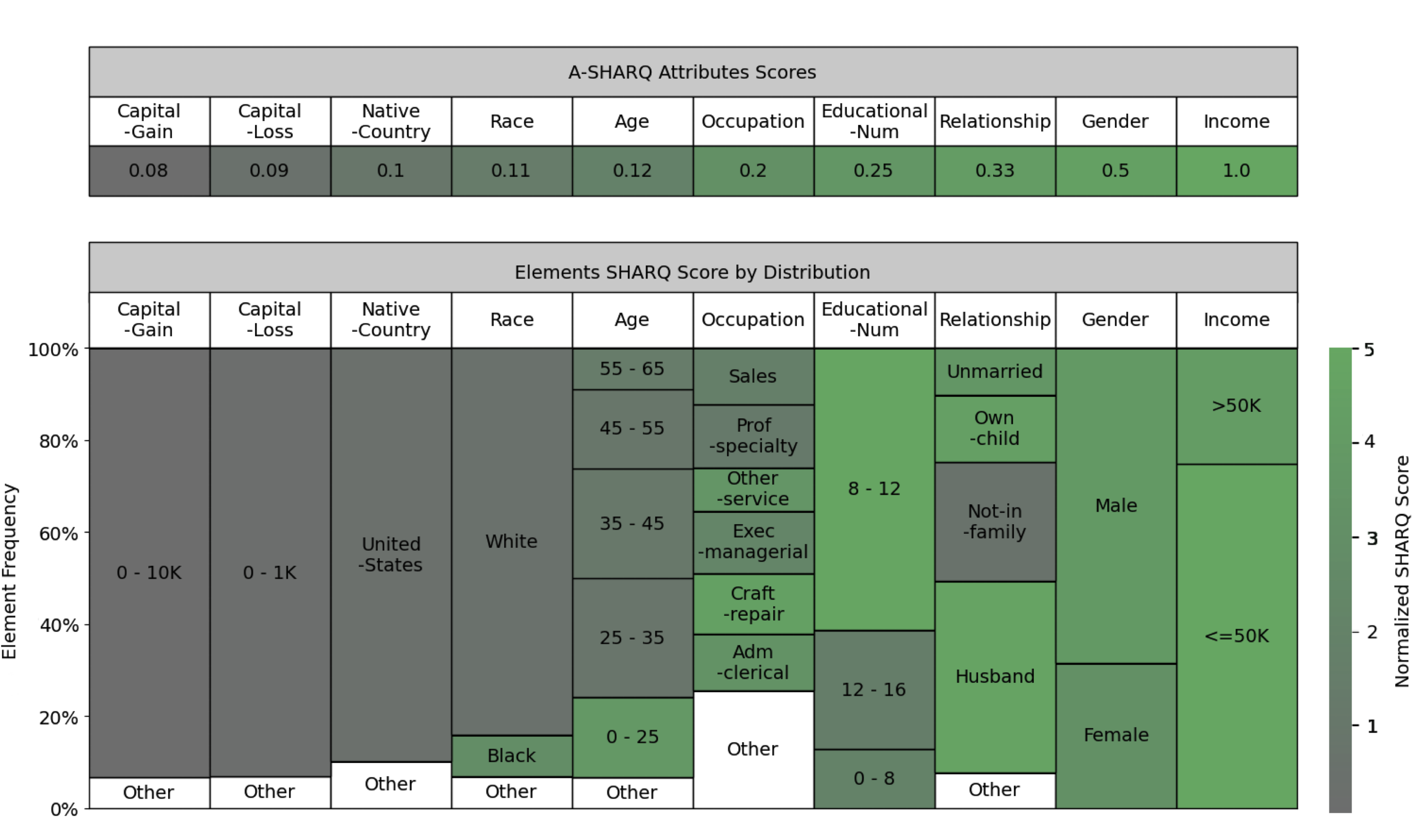}
        \label{4figs-a}
    \vspace{-4mm}
    \caption{A-\system{} attribute importance scores}
    \label{fig:att_importance}
\end{figure*}

Our analysis above reveals some highly-frequent elements with nearly zero \system{} score. 
We next show how to detect such elements with a \textit{normalized} \system{} score. Then, using the normalized \system{}, we define the \system{} score for \textit{rules} and use it to identify important, non-redundant rules.

\paragraph*{Normalized \system{}} 
To provide a normalized \system{} score for an element $e \in E$ which accounts for the element's frequency, we use the following: First, let $R_{SHARQ}(e)$ be the \textit{rank} of element $e$ w.r.t. the score $\naivesingle$ (a rank of $i$ is given to the top-$i$ scoring elements). 
%(i.e., $R_{SHARQ}(e)=1 \iff \naivesingle = \max_{e'\in E} SHARQ_{(E,R)}(e')$).
We also denote by  $R_{freq}(e)$ the \textit{frequency} rank of an element $e$, s.t., the most frequent element is ranked 1. Then the normalized score is given by:
$$\widehat{\system{}}_{(E,R)}(e) = \frac{R_{freq}(e)}{R_{SHARQ}(e)}$$
%$\widehat{\system{}}_{(E,R)}(e)$ is defined by %$\frac{R_{freq}(e)}{R_{SHARQ}(e)}$. 
Intuitively, the \textit{lower} the normalized \system{} score the more \textit{redundant} the element $e$ is. 
%\vspace{1mm}

Using normalized \system{}, we define the \textit{rule-level} \system{}:
$$\rulescoresingle = \min_{e \in \mathcal{E}(r)} \widehat{SHARQ}_{(E,R)}(e)$$ 
\noindent
Namely, the \system{} rule score is determined by its lowest-scoring element.

Figure~\ref{fig:alt_rules} depicts several rules alongside their IS interestingness scores and corresponding R-\system{} score. 
For each rule, we colored the element obtaining the lowest normalized \system{} score. The color is proportional to the score -- grey symbolizes low scores, and green symbolizes high scores.  
%Rule elements in the LHS or RHS with particularly low R-\system{} scores are colored in grey, while those with high scores are colored in green.

For example, Rule \#1, (\textit{educational-num},\textit{8-12}),  (\textit{Native-country}, \textit{United-States}) $\longrightarrow$ (\textit{income},\textit{$<=50K$}) obtains a high IS score of 0.7, yet a low R-\system{} score of 0.21. This difference can be explained by the fact that the rule contains the element (\textit{Native-country}, \textit{United-States}), which has a low normalized \system{} score (colored in light grey). An equivalent rule exists \textit{without} this element (Rule \#2): (\textit{educational-num},\textit{8-12}) $\longrightarrow$ (\textit{income},\textit{$<=50K$}). Rule\#2 has an even higher IS score (0.73), and a substantially higher R-\system{} score of 4. 

In a similar manner, Rule \#3, (\textit{Gender},\textit{Male})$\longrightarrow$ (\textit{Capital-Gain}, \textit{0-10K}),  (\textit{Relationship}, \textit{Husband})  obtains a low R-\system{} score of $0.06$.  When eliminating the element (\textit{Capital-gain},\textit{0-10K}) we obtain Rule \#4, (\textit{Gender}, \textit{Male})$\longrightarrow$ (\textit{Relationship}, 
 \textit{Husband}), which obtains a higher IS score of 0.77 and a higher R-\system{} score of 1.75. 

Note that when setting a rule-level \system{} threshold of 0.21 (the R-\system{} score of Rule\#1), a total of 68250 (95\%) rules fall below the threshold. Eliminating these rules allow the user to focus on a much smaller set of 3140 important, non-redundant rules.

%Using the rule-level \system{} score one can reduce the number of rules and focus only on important, non-redundant ones.  For example, in our example rules set, when setting the rule-level \system{} threshold of \amit{YY}, we obtain that 74341 rules (88\%) fall below the threshold, leaving 

%After calculating the normalized \system{} scores for all elements, 
%We then examined the rules subset that is only composed of elements with above-threshold normalized \system{} score, namely:  $$R^+=\{ r \in R \wedge \forall e \in \mathcal{E}(r), \widehat{\system{}}_{(E,R)}(e) > \theta  \}$$ 

\subsection{Rules-driven Attribute Importance}
\label{ssec:app_feature_importance}
%Leveraging the individual \system{} scores, w
We next show another common explainability use case, attribute (feature) importance, in which we use \system{} to detect the most significant attributes for the given set of association rules. 

%\susan{What is $\mathcal{E}_a$?  Do you mean $\mathcal{E}_a(D)$ (in which case isn't this just $\mathcal{A}$) or $\mathcal{E}_a(R)$ or $\mathcal{E}_a(E)$?}\amit{see comment above}

To do so, we define an \textit{attribute}-level \system{} score: $$\attrscoresingle = \frac{\sum_{e \in E_a \cap \mathcal{E}(R)} \widehat{SHARQ}_{(E,R)}(e) }{E_a}$$ 
\noindent
\fix{where $E_a$ is the subset of $E$ whose elements have attribute $a$.}
This is the mean \textit{normalized} \system{} score for the elements of attribute $a$ that appear in at least one rule. 

 Figure~\ref{fig:att_importance} depicts the importance scores of sample attributes in the Adult dataset, sorted from left (lowest) to right (highest). Below the attribute importance scores, the figure shows the attribute elements. The cell background colors reflect the \textit{element}-level \system{} scores, from grey (low) to green (high), and the cell size reflects the frequency of the element with respect to the attribute value distribution (namely, longer cells represent more frequent elements).
%\susan{Need to say what the color means, what is good and what is bad.}

Notice that the \textit{Income} and \textit{Gender} columns obtain the highest attribute scores, as both contain only high-importance individual elements. The \textit{Relationship} column is ranked third, as one of its elements, \textit{'Not-in-Family'}, has a low importance score (colored in grey in the figure).

Also note that the columns \textit{Capital-loss}, \textit{Capital-gain}, \textit{Native-country}, and \textit{Race} obtain the lowest A-\system{} importance scores.  This is because they all contain a single dominant element with a high frequency, yet low importance (see their corresponding element-level \system{} scores in the lower part of Figure~\ref{fig:att_importance}).

The attributes importance scores can be used, e.g., to reduce the dimensionality of the data (by disregarding low-scoring attributes) when performing analytical tasks.

%\susan{Need a closing thought here.  E.g. "This could be used by the analyst to reduce the dimensionality of the data for downstream analysis."  ???} \amit{One can eleimnate such columns for dimnesionality reduction etc.}

\paragraph*{Summary \& Discussion}
In this section, we gave two additional explainability applications based on the element-level \system{} scores: rule-level and attribute-level importance scores. As demonstrated in~\cite{lundberg2020local}, a similar analysis in the context of ML explainability is performed using the SHAP~\cite{shap_ml} explainability framework, where individual SHAP scores are aggregated over multiple predictions. Using such aggregations, one can explain global aspects of the ML model, including global feature importance scores and feature dependency analysis. In our context, additional interesting aggregations of the element-level \system{} scores include feature interaction, which examines the change in importance of one feature (element) as a function of the values of a second feature; calculation of element \textit{subsets} importance; and \system{} correlations.

%% file: experiments.tex
\section{Experimental evaluation}
\label{sec: exp}

We conducted our experiments on a evaluation set consisting of $45$ different rule mining results,
generated using the Apriori algorithm~\cite{agrawal1994fast} with different settings and datasets.
The evaluation setup is discussed in Section~\ref{ss: setup}. 
We then describe four different sets of experiments.

First, we examine the performance of our optimized \sysopt{} algorithm and compare it to the naive calculation of the \system{} formula. In line with our theoretical cost analysis, the results in Section~\ref{ss: opt-naive} show that the naive calculation is infeasible as it iterates over more than 65138729 coalitions on average, whereas \sysopt{} uses only 8524.

Next, we analyze the performance of our multi-element \sysopt{} algorithm in order to see whether it indeed provides a significant performance improvement compared to a sequential calculation of \sysopt{}.  As we show in Section~\ref{ss: all-elements}, the multi-element algorithm provides an average of \rev{13.8X} improvement in running time.

We then test the ability of alternative approaches such as \textit{influence} and \textit{$I_{TOP}$} (see Section~\ref{ssec:defs}) to approximate the score of \system{}. Our findings, provided in Section~\ref{ss: baselines}, show that a direct approximation for \system{} based on ~\cite{shap_ml} is significantly better at preserving the original \system{} elements ranking while maintaining similar running times to \textit{Influence} and $I_{TOP}$.

Finally, in Section~\ref{ssec:ablation} we examine the effect of using different notions of interestingness for rules and rule sets. We find that the \system{} scores are robust and consistent, and do not significantly change when using such alternative notions. 

\subsection{Experimental Setup \& Evaluation Set}
\label{ss: setup}
We first describe our experimental setup and explain how we constructed our evaluation set, consisting of \rev{67} different rule mining settings applied on four underlying datasets. 

\vspace{2mm}
\noindent\textbf{Implementation and Default Configuration.} \system{} is implemented in Python 3.10. It uses an Apriori algorithm implementation\footnote{\url{https://efficient-apriori.readthedocs.io/en/latest/}} to generate the rules, and  Pandas~\cite{pandas} to store and manipulate the underlying data. The experiments were run on a Windows 11 laptop with 48GB RAM and 2200 Mhz 12 cores i7 processor.

As default implementation choices for \system{}, we use the IS score as the rules interestingness measure, and the \textit{max} aggregation for rules set: $I(R)\coloneq max_{r\in R} IS(r)$.
In Section~\ref{ssec:ablation} we discuss alternative scoring and aggregations.

\vspace{2mm}
\noindent\textbf{Evaluation Set Construction.}
We constructed an evaluation set consisting of 45 different rule mining results with diverse characteristics. 
Each instance in our evaluation set is a pair $\langle \df, \rulesdf \rangle$, containing a dataset and the set of rules extracted from it. 
We used the following underlying datasets to generate the rules: \textbf{ 
 Adults}~\cite{adults_dataset}, containing 49K rows and 16 columns, \textbf{Spotify songs}~\cite{spotify_dataset}
, containing 174K rows and 22 columns, 
\textbf{Flight Delays}~\cite{flights_dataset}, containing 5.8M rows and 30 columns, \textbf{ Isolet}~\cite{isolet_dataset}, containing 7.8K rows and 617 columns, \rev{\textbf{Covid-19}\cite{covid19_dataset}, containing 316.8K rows and 27 columns, and \textbf{Adult-ACS}~\cite{adults_acs_dataset}, containing 144.2K rows and 828 columns.} 

We then used multiple rule mining pipelines on each dataset that included the following steps: (1) sampling the rows, (2) binning numeric columns, (3) running a rule mining tool with a predefined support threshold, and (4) filtering the resulted rules according to their lift score~\cite{brin1997beyond}.

Note that steps (1) and (2) are necessary for a rule mining algorithm to work properly in a reasonable amount of time. 
We discarded configurations that took longer than 8 hours to run.

We varied the sample size between 2K and 10K, the number of bins between 3 and 5, the rules support threshold between 0.05 and 0.2, and the lift threshold between 1.05 and 1.2.\footnote{For lift scores showing negative association ($\leq 1$), we use $\frac{1}{score}$ before matching with the threshold.} 
%0.05 and 0.2 for rules with lift score higher than 1 and 5 to 20 for rules with lift score lower than 1. \amit{change}

Our rule mining pipelines generated a total of \rev{67} distinct rules sets. Table \ref{tab:datasets} details the characteristics of the rules generated for each dataset in terms of ranges (minimum and maximum) of the number of rules, average rule length, and the number of elements participating in rules ($\mathcal{E}(R)$). %In Sections~\ref{ssec: single}-\ref{ss: baselines} we examine the effect of these rule sets properties on the performance of our algorithms. 

\begin{table}[t]
\centering
\small
%\ttfamily
    \begin{tabularx}{\columnwidth}{|X| c !{\vrule width 1.2pt} c| c| c|}   
    \hline 
    \textbf{Dataset} & \makecell{\textbf{Num. of} \\ \textbf{Rule Sets}} & \makecell{\textbf{Num. of} \\ \textbf{Rules}} & \makecell{\textbf{Avg.} \\ \textbf{Rule Len.}} & \makecell{\textbf{Num. of} \\ \textbf{Elements}}  \\
    \hline \noalign{\hrule height 0.9pt}
    $Adults$ & 14 & [563, 471474] & [3.8, 6.8] & [21, 34] \\
    \hline
    $Flights$ & 14 & [216, 26964] & [3.1, 5.2] & [19, 40] \\
    \hline
    $Isolet$ & 3 & [204, 498834] & [3.1, 6.7] & [27, 40] \\
    \hline
    $Spotify$ & 14 & [67, 16867] & [2.6, 4.7] & [17, 40] \\
    \hline
    \rev{$Covid\-19$} & \rev{11} & \rev{[58011, 185214]} & \rev{[5.34, 5.38]}
 & \rev{[37, 38]} \\
    \hline
    \rev{$Adult\text{-}ACS$} & \rev{11} & \rev{[345, 421487]} & \rev{[4.5, 7.2]} & \rev{[12, 22]} \\
    \hline
    \end{tabularx}
\captionof{table}{\rev{Properties of generated rule sets [min, max]}}
\vspace{-2mm}
\label{tab:datasets}
\end{table}

\begin{figure*}[t!]
\vspace{-5mm}
    \centering
    \begin{subfigure}{1\linewidth}
        \centering
        \includegraphics[scale=0.4]{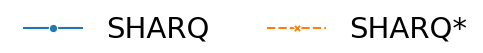}
        \vspace{-6mm}
        \caption*{} 
    \end{subfigure}
    \vspace{2mm} % Add some vertical space between the legend and the subfigures
    \begin{subfigure}[b]{0.25\linewidth} 
        \centering
        \includegraphics[scale=0.25]{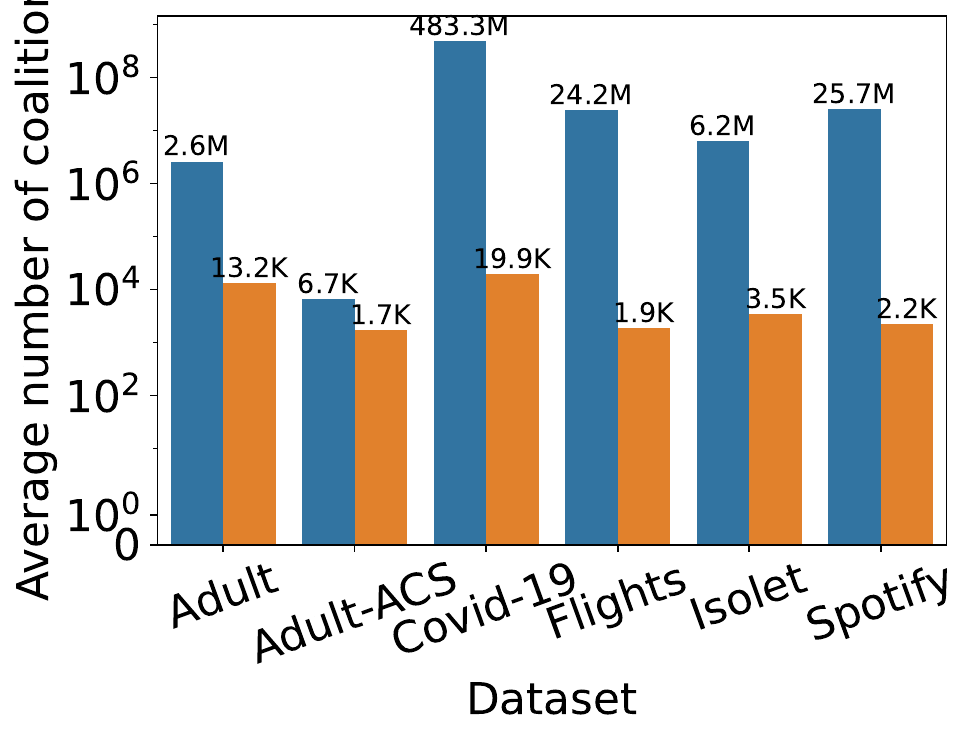}%
        \caption{\rev{\# coalitions by dataset }}
        \label{sfig:single_by_dataset} 
    \end{subfigure}%
    \begin{subfigure}[b]{0.25\linewidth} 
        \centering
        \includegraphics[scale=0.25]{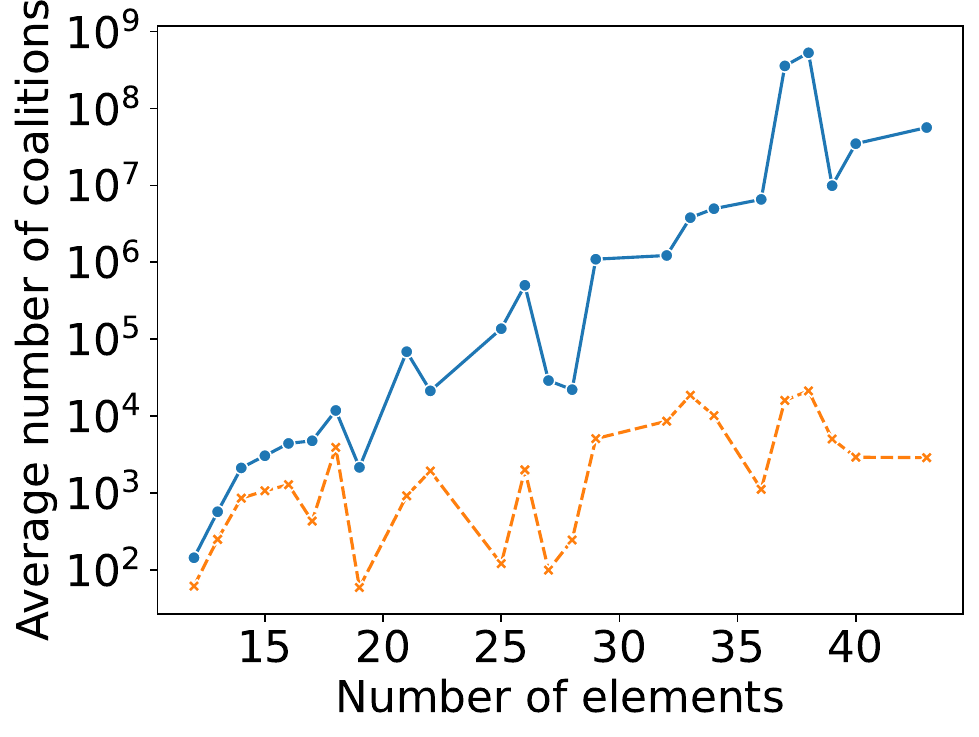}%
        \caption{\rev{\# coalitions by \# elements }}
        \label{sfig:single_elements_num}
    \end{subfigure}%
    \begin{subfigure}[b]{0.25\linewidth} 
        \centering
        \includegraphics[scale=0.25]{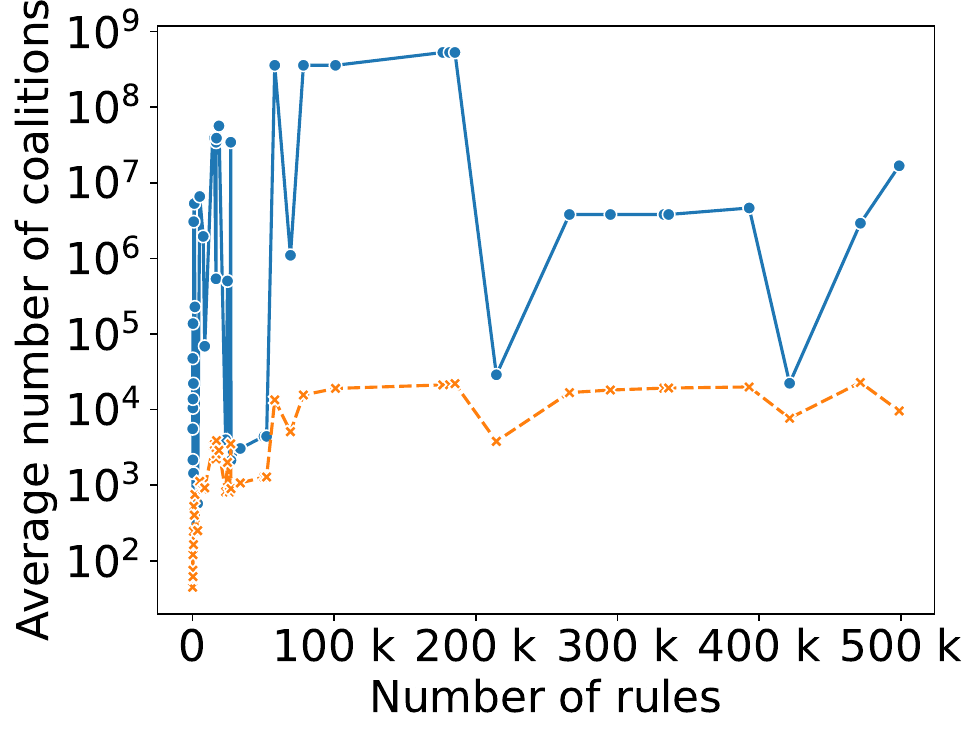}%
        \caption{\rev{\# coalitions by \# rules}}
        \label{sfig:single_num_of_rules} 
    \end{subfigure}%
    \begin{subfigure}[b]{0.25\linewidth} 
        \centering
        \includegraphics[scale=0.25]{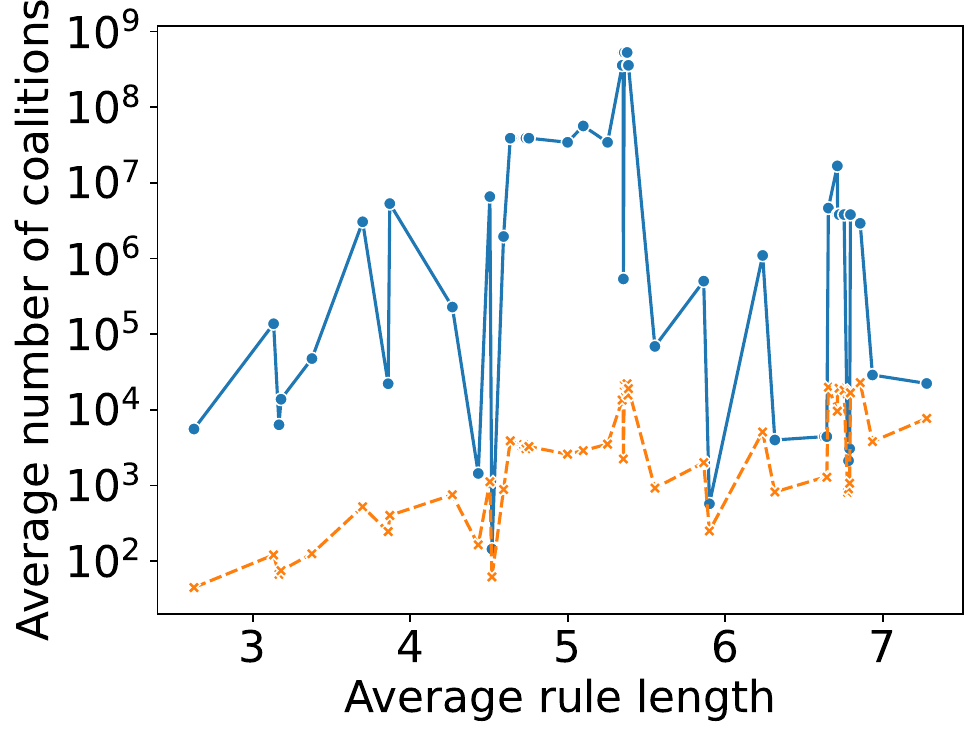}
        \caption{\rev{\# coalitions by avg. rule length}}
        \label{sfig:single_rule_len}
    \end{subfigure}%
    \vspace{-2mm}
    \caption{\rev{Comparison of naive and optimized approaches }}
    \label{fig:coalition_num_exsp}
\end{figure*}

\begin{figure*}[t!]
\vspace{2mm}
    \centering
    \begin{subfigure}{1\linewidth}
        \centering
        \includegraphics[scale=0.6]{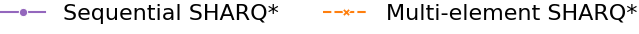}
        \vspace{-6mm}
        \caption*{} 
        \label{fig:user_study_legend}
    \end{subfigure}
    \vspace{2mm} % Add some vertical space between the legend and the subfigures
    \begin{subfigure}[b]{0.25\linewidth} 
        \centering
        \includegraphics[scale=0.25]{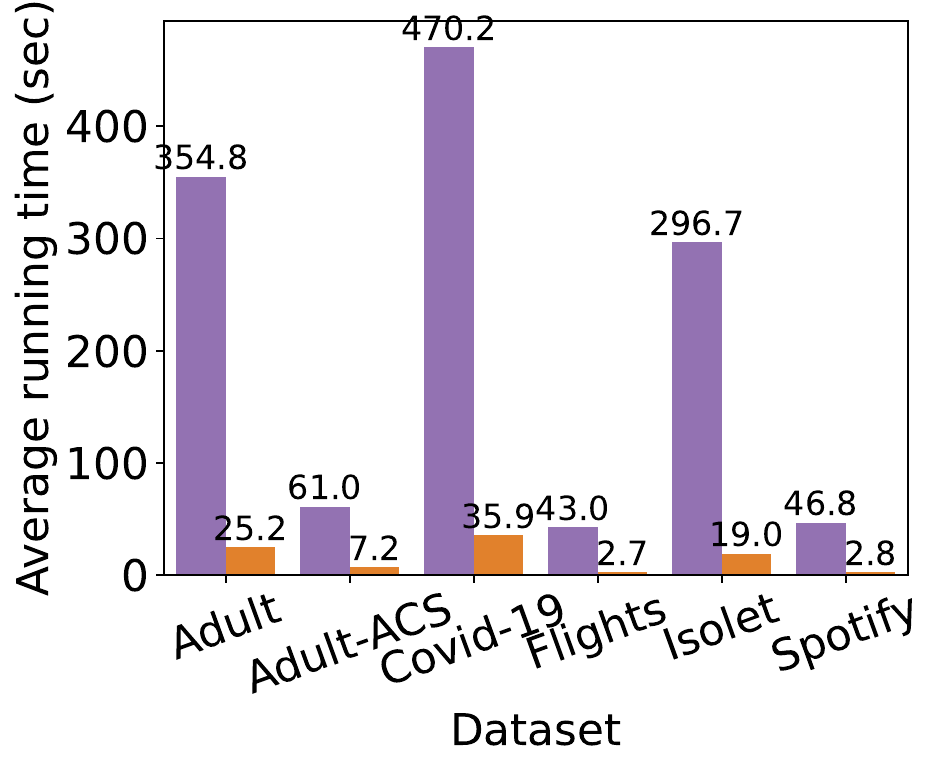}%
        \caption{\rev{Avg. time by dataset}}
        \label{sfig:seq_by_dataset} 
    \end{subfigure}%
    \begin{subfigure}[b]{0.25\linewidth} 
        \centering
        \includegraphics[scale=0.25]{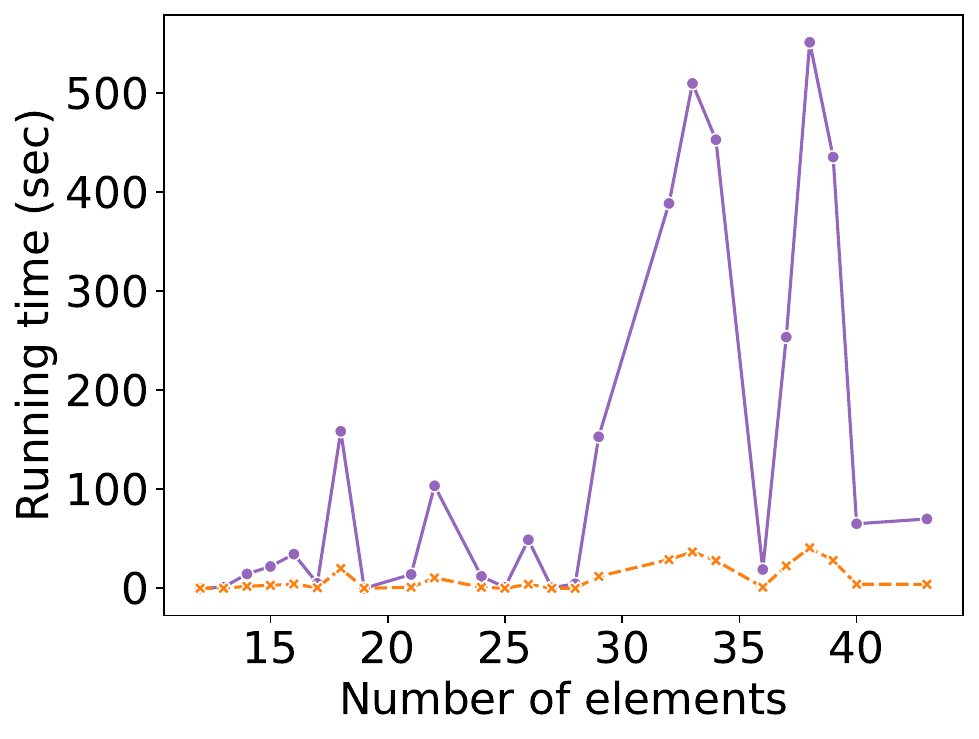}%
        \caption{\rev{Avg. time by \# elements}}
        \label{sfig:seq_elements_num}
    \end{subfigure}%
    \begin{subfigure}[b]{0.25\linewidth} 
        \centering
        \includegraphics[scale=0.25]{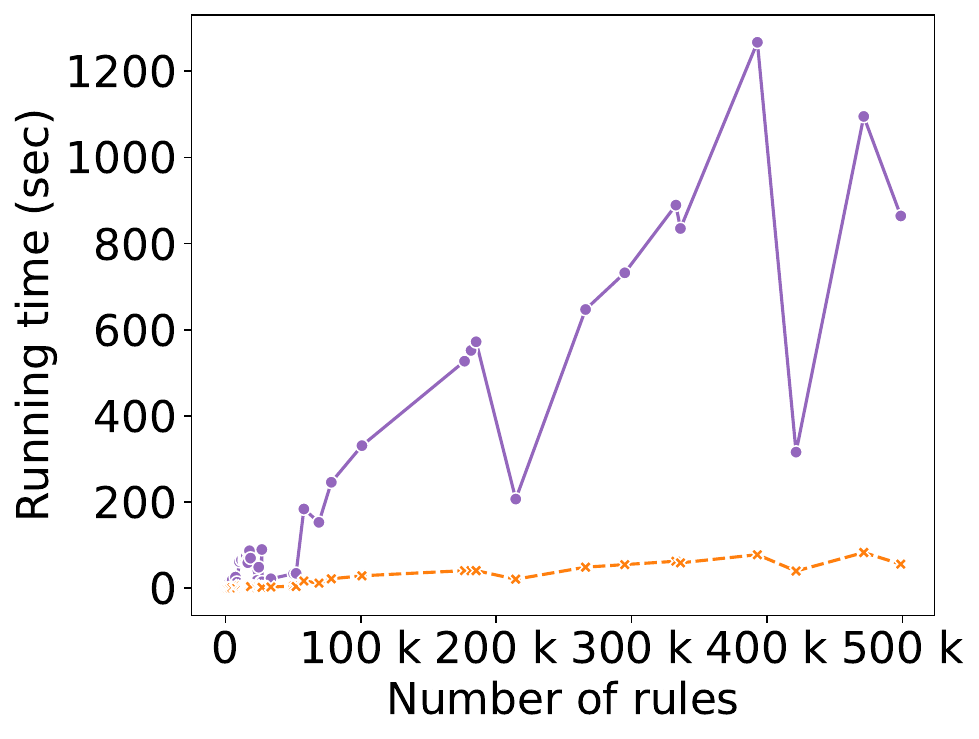}%
        \caption{\rev{Avg. time by rule set size}}
        \label{sfig:seq_num_of_rules} 
    \end{subfigure}%
    \begin{subfigure}[b]{0.25\linewidth} 
        \centering
        \includegraphics[scale=0.25]{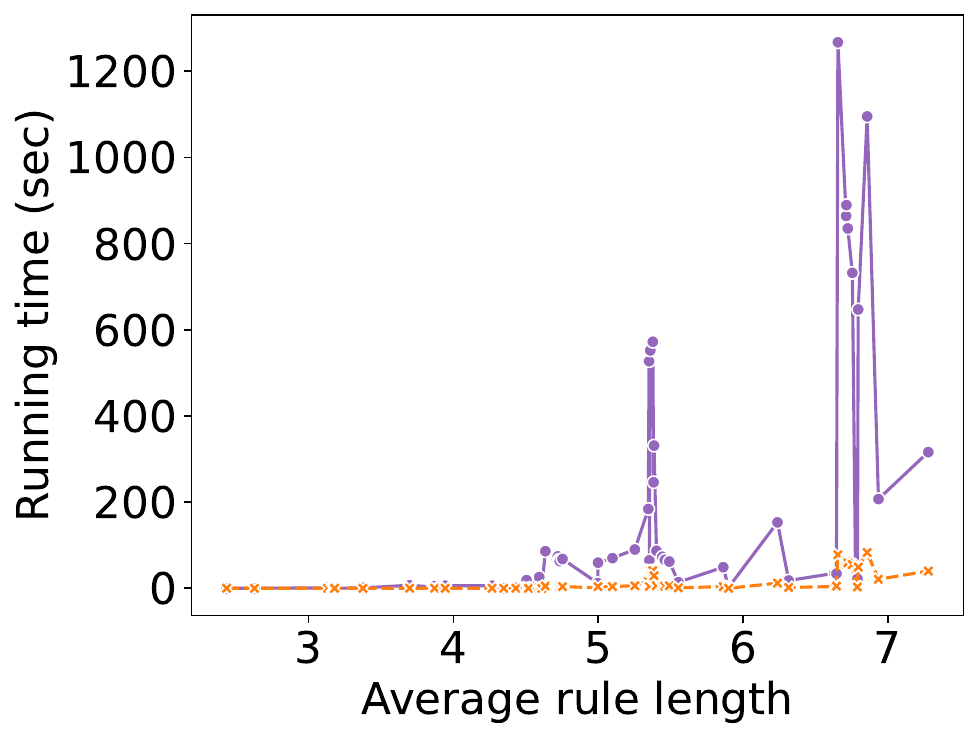}
        \caption{\rev{Avg. time by avg. rule length}}
        \label{sfig:seq_rule_len}
    \end{subfigure}%
    \vspace{-2mm}
    \caption{\rev{Comparison between multi-element and sequential approaches}}
    \label{fig:sequential_exsp}
\end{figure*}

\subsection{Single Element \sysopt{} Analysis }
\label{ss: opt-naive}
We first gauge the effectiveness of our single element \sysopt{} algorithm
compared to the naive calculation of \system{}, which instantiates all valid coalitions $\mathcal{C}(e,E)$ as explained in Section~\ref{ssec: formula}. Recall that by definition $\naivesingle = \optsyssingle$, therefore we only examine the efficiency of calculations. 

Since the running times and memory consumption of the naive \system{} calculation were extremely high (taking more than 8 hours for small rules sets with under 1K rules, and up to several days and even more for larger ones.)  
we compare the number of elements coalitions considered in the calculation rather than running times. As explained in Section~\ref{ssec: formula}, this is the main factor affecting the running times. 

In this series of experiments, we calculated the number of coalitions used by $\naivesingle$ and by $\optsyssingle$, i.e. $|\validcoalitions|$ and $|\mathcal{C}^*(e,E)|$ respectively.
This was done for each element $e \in E$ in each evaluation instance $\langle \df, \rulesdf \rangle$, where $E=\mathcal{E}(R)$ (a total of 1531 executions). 
Rather than materializing $|\validcoalitions|$, we used the calculation from our analysis in Section~\ref{ssec: formula}: $|\validcoalitions| = \prod_{a\in \{{attr(E)}\} -\{attr(e)\}} (|\elemsshort_\column| + 1)$. 

%The sizes of the coalitions were determined by first counting the unique values in each attribute. Then, for each attribute, we calculated the number of ways to form a coalition from elements that do not share the same attribute. This was done by multiplying the number of possible elements by the number of options for not participating in the coalition (i.e., the number of possible elements plus one).

The results are depicted in Figure~\ref{fig:coalition_num_exsp}; note that the y-axis in all sub-figures use a logarithmic scale.
Figure~\ref{sfig:single_by_dataset} shows the average number of coalitions for a single element $e$, grouped by the \textit{underlying dataset}. 
%$D \in \{\text{Adult}, \text{Flights}, \text{Isolet}, \text{Spotify}\}$. 
The most substantial difference is observed in the \rev{\textit{Covid-19}} dataset, where the \sysopt{} calculation generated \rev{19.9K} coalitions whereas the naive approach generated \rev{483.3M} coalitions. The smallest difference occurred in the \rev{\textit{Adult-ACS}} dataset, where the optimized calculation generated \rev{1.7K} coalitions whereas the naive approach generated  \rev{6.7M} coalitions, on average.

We then investigate how the properties of the mined rule sets affect our computation.
Figure~\ref{sfig:single_elements_num} shows the average number of coalitions as a function of the number of distinct elements appearing in the rule set. 
As expected, the number of coalition for the naive \system{} increases exponentially. This is due to the fact that coalitions are formed from every possible subset of elements, excluding those containing more than one element with the same attribute (as defined in Section~\ref{ssec: formula}). In contrast, \sysopt demonstrates a much slower growth. This is because the optimized set of coalitions $\optsyssingle$ is based on the number and size of rules rather than by the number of distinct elements.  While in theory the number of rules can be exponential in the number of elements, in practice it is much smaller.

Next, Figure~\ref{sfig:single_num_of_rules} shows the average number of coalitions as a function of the number of rules in $\rulesdf$. 
For both the naive \system{} and \sysopt{} we observe a slight increase in the number of coalitions as the rules set size increases.

Figure~\ref{sfig:single_rule_len} shows the average number of coalitions as a function of the average rule length, $|r|$, i.e. the number of elements contained in a rule.  
As expected, for both algorithms, the number of coalitions increases as the rules contain more elements, yet the naive approach utilizes at least three orders of magnitude more coalitions. 

As for running times, calculating the \system{} score for an element using \sysopt{} takes \rev{6.6} seconds, on average, compared to hours and even more if using the naive approach that considers all valid element coalitions (we stopped the computation process after 8 hours of execution).

\subsection{Multi-element vs Sequential \sysopt{}}
\label{ss: all-elements}
In the multi-element setting, we want to calculate  \system{} scores for all elements in a set $E$.
In this experiment, we therefore investigate the performance of our multi-element \sysopt{} algorithm (Algorithm~\ref{algo:multi}) compared to a \textit{sequential} execution of the single element \sysopt{} (Algorithm~\ref{algo:single}), computed for each $e \in E$, where $E=\mathcal{E}(R)$.  This time, as both approaches utilizes the optimized \sysopt{} calculation, we can compute actual running times rather than the number of coalitions as done for the naive \system{} performance evaluation.

Figure~\ref{fig:sequential_exsp} shows the running time for calculating \system{} scores for all elements in $\mathcal{E}(R)$, measured across all of our 45 $\langle \df, \rulesdf \rangle$ evaluation instances.
First, in Figure~\ref{sfig:seq_by_dataset}, we show the running times of the two solutions, averaged over all rule sets per dataset (i.e., Adult, Flights, Isolet, and Spotify). Observe that the multi-element algorithm achieves a significant improvement over the sequential approach, on average reducing running times by \rev{13.6X} (Adult dataset), \rev{17.5X} (Flights), \rev{20.4X }(Isolet), \rev{16.2X} (Spotify),\rev{ 8X (Adult-ACS) and 12.6X (Covid-19}). The average reduction across all \rev{67} instances is \rev{13.8X}. 
Recall from Section~\ref{ssec: alg} that the minimal reduction factor of Algorithm~\ref{algo:multi}, the multi-element algorithm, is approximately $\tau$ (i.e., the maximal rule length). For the Adult, \rev{Covid-19}, Flights, and Isolet datasets, the maximal rule size is 8, and for the Spotify \rev{and Adult-ACS} datasets, the maximal rule size is 7. Across all \rev{67} instances, the \textit{minimal} improvement achieved by Algorithm~\ref{algo:multi} is \rev{6.7X}.

To complete the picture, Figures~\ref{sfig:seq_elements_num}-\ref{sfig:seq_rule_len} show the running time as a function of the number of elements, the size of the rule set, and the average rule length, respectively. We observe an apparent linear difference between these approaches, especially for complex rule sets with more than 30 elements, 20K rules, and when the average rule length is above 4.5. 
%When computed on rule sets with lower values for these properties, the multi-element algorithm is slightly less effective. However, even in these cases, the \textit{minimal} improvement achieved by Algorithm~\ref{algo:multi} is 7.4X.

%\usepackage[table,xcdraw]{xcolor} % Include this in the preamble

\begin{table}[t]
\centering
\small

\begin{tabularx}{\columnwidth}{|X| c| c| c| c|}
    \hline
    \rev{\textbf{Baseline}} & \textbf{p@10} & \textbf{ap@10} & \textbf{Rank Corr.} & \textbf{Run Time} \\
    \hline
    %\rowcolor{gray!40}
     \multicolumn{5}{|c|}{\textbf{Alternative Measures for Contribution}} \\
    \hline
    $Influence$ & \rev{0.71} & \rev{0.51} & \rev{0.76} & \rev{10.14s} \\
    \hline
    $I_{TOP}$ & \rev{0.73} & \rev{0.51} & \rev{0.67} & \rev{10.48s} \\
    \hline
     \multicolumn{5}{|c|}{\textbf{Approximated \system{} Calculation}} \\
    \hline
    Kernel-Weighting & \rev{0.92} & \rev{0.84} & \rev{0.93} & \rev{15.46s} \\
    \hline
    \rev{Sobol-Sequence} & \rev{0.75} & \rev{0.65} & \rev{0.85} & \rev{13.38s} \\
    \hline
    \rev{Stratified Sampling} & \rev{0.74} & \rev{0.66} & \rev{0.85} & \rev{19.38s} \\
    \hline
    \rev{Monte-Carlo Antithetic} & \rev{0.73} & \rev{0.61} & \rev{0.83} & \rev{27.82s} \\
    \hline
    \rev{Monte Carlo} & \rev{0.76} & \rev{0.64} & \rev{0.82} & \rev{12.85s} \\

    \hline
\end{tabularx}
\captionof{table}{\rev{\system{} approximation quality and running times}}
\vspace{-4mm}
\label{tab:approximations}
\end{table}

%\subsection{Alternative measures vs. \system{} Approximation}
\subsection{Approximations \rev{and Alternative Approaches}}
\label{ss: baselines}
In Section~\ref{ssec:defs}, we described two possible alternative methods for calculating the contribution of an element $e$ to the interestingness of a rules set $R$, $I_{TOP(r)}(e)$ and $Influence_{(R)}(e)$. While both methods can be computed more efficiently than \system{} scores, we showed in Example~\ref{exa:interestingness} that they fail to adequately capture the difference in the contribution of elements. 

In the next set of experiments, our goal is to validate this observation and examine whether $I_{TOP}$ and $influence$ can provide a good \textit{approximation} for the \system{}.
Since both alternative approaches use different calculation methods, \rev{we expect their output to be different than \system{}. We therefore calculate the following metrics, in order to determine how their ranking and top-scoring elements compare to those returned by the \system{} calculation:}
\textit{(1) p@10}, which compares the top-10 \system{} elements to the top-10 w.r.t. the baseline score and then calculates $\frac{\text{\#top-10 matches}}{10}$, i.e., the number of matching elements in the two top-10 lists, divided by 10; \textit{(2) ap@10},  defined as $\frac{1}{10}\sum_{k=1}^{10}{p@k}$, namely the average p@k score from 1 to 10 (p@k generalizes the p@10 measure defined above, for arbitrary values of $k$); and \textit{(3) Spearman Rank Correlation} which measures the correlation between the \textit{rank order} of the baseline scores compared to the \system{} scores. We also report the running times of each baseline for computing the scores of all $\mathcal{E}(R)$ elements.

\rev{We compare the alternative contribution calculation methods to five direct \system{} approximations, previously suggested for Shapley-based calculations~\cite{shap_ml,mitchell2022sampling}. } The first is (1) \textit{Shapley kernel weighting}, based on~\cite{shap_ml}, in which \fix{permutations} are sampled after assigned a weight of $\frac{|E|-1}{\binom{|E|}{|S|}|S|(|E|-|S|)}$. \rev{The remaining four are permutation sampling techniques suggested in the context of Shapley approximations~\cite{mitchell2022sampling}:  (2) \textit{Monte-Carlo}, in which the permutations are sampled i.i.d.;  (3) \textit{Antithetic Monte-Carlo}, where half of the permutations are sampled uniformly, and for each randomly-sampled permutation we also take its complement; (4) \textit{Stratified sampling,} another common technique for reducing variance, where we first segment the permutations by size then take a uniform sample from each group; and (5) \textit{Sobol-Sequence Sampling}, which provides a better coverage of the sampled domain by avoiding resampling similar points.}
%To examine their utility, we compare their approximation quality and running times against a \textit{direct} \system{} approximation, based on Shapley kernel weighting~\cite{shap_ml}. This is a generic sampling method for all Shapley based calculations, in which coalitions are sampled after assigned a weight of $\frac{|E|-1}{\binom{|E|}{|S|}|S|(|E|-|S|)}$.  

%Since both alternative approaches use different calculation methods, rather than comparing the exact \system{} to the approximated ones we used the following evaluation metrics: \textit{(1) p@10}, which compares the top-10 \system{} elements to the top-10 w.r.t. the baseline score and then calculates $\frac{\text{\#top-10 matches}}{10}$, i.e., the number of matching elements in the two top-10 lists, divided by 10; \textit{(2) ap@10},  defined as $\frac{1}{10}\sum_{k=1}^{10}{p@k}$, namely the average p@k score from 1 to 10 (p@k generalizes the p@10 measure defined above, for arbitrary values of $k$); and \textit{(3) Spearman Rank Correlation} which measures the correlation between the \textit{rank order} of the baseline scores compared to the \system{} scores. We also report the running times of each baseline for computing the scores of all $\mathcal{E}(R)$ elements.

%As for running time, for each approach we measure the \textit{time fraction}, calculated as $\frac{\text{baseline time}}{\text{\sysopt{} time}}$ (computed on \textit{all} elements, using Algorithm~\ref{algo:multi}). \amit{update me }

\rev{Table~\ref{tab:approximations} first presents the approximation performance of the alternative contribution calculation approaches: $I_{TOP}$ and Influence, and then gives the results for the direct \system{} approximations.
Observe that all direct \system{} approximation outperforms both alternative approaches across all approximation quality metrics. Notably, the Shapely kernel weighting method obtains the highest approximation quality, achieving scores of 0.92, 0.84, and 0.93 for p@10, ap@10, and rank correlation, with an overall average running time of 15.46s. In comparison, the alternative contribution methods are faster (10.14s and 10.48s) but have an inferior quality, not exceeding 0.73, 0.51 and 0.76 for p@10, ap@10, and rank correlation. 
Note that the Sobol-sequence and Monte-carlo approaches obtains better results than $I_{TOP}$ and $Influence$, with almost the same running times (13.38s and 12.85s) respectively. 

%presents the differences between the direct \system{} approximation and the alternative approaches, $I_{TOP}$ and Influence. Observe that the direct \system{} approximation significantly outperforms both alternative approaches across all approximation quality metrics, achieving scores of 0.89, 0.8, and 0.92 for p@10, ap@10, and rank correlation, respectively. In comparison, the runner-up $I_{TOP}$ scores are 0.78, 0.53, and 0.81, respectively.
%In terms of running time, the direct \system{} approximation takes less than a second longer than the alternative approaches: 18.36s compared to 17.51s and 17.98s reported by $I_{TOP}$ and Influence.

These findings suggest that the alternative approaches are \textit{not} good estimators for \system{} scores, as the direct \system{} approximations maintain higher accuracy and comparable running times.
}

%\susan{Minor rewording below, you might want to check.}
\rev{
In addition to these results which show the deviation in element rankings between \system{} and alternative contribution scores, we have empirically observed (e.g. in Example~\ref{exa:interestingness}) the inadequacy of $I_{TOP}$ and $influence$ in capturing importance differences between elements. However, a more comprehensive analysis should be conducted using qualitative comparisons, such as case studies, user studies, and quantitative evaluations. This approach has recently been proposed for evaluating and comparing established XAI methods in the context of supervised learning models~\cite{salih2024perspective} and autonomous agents~\cite{rosenfeld2021better}. 

Since \system{} is, to the best of our knowledge, the first framework for quantifying the importance of elements in an association rule set, we defer the development of such evaluation methods to future work.}

%Compared to the exact \sysopt{} running times, the approximation approaches reduce 38\% (Direct), 45\% ($I_{TOP})$ and 42\% ($Influence$). 

\begin{figure}[t!]
    \centering
    \begin{subfigure}[b]{0.48\columnwidth} 
        \centering
        \includegraphics[width=\linewidth]{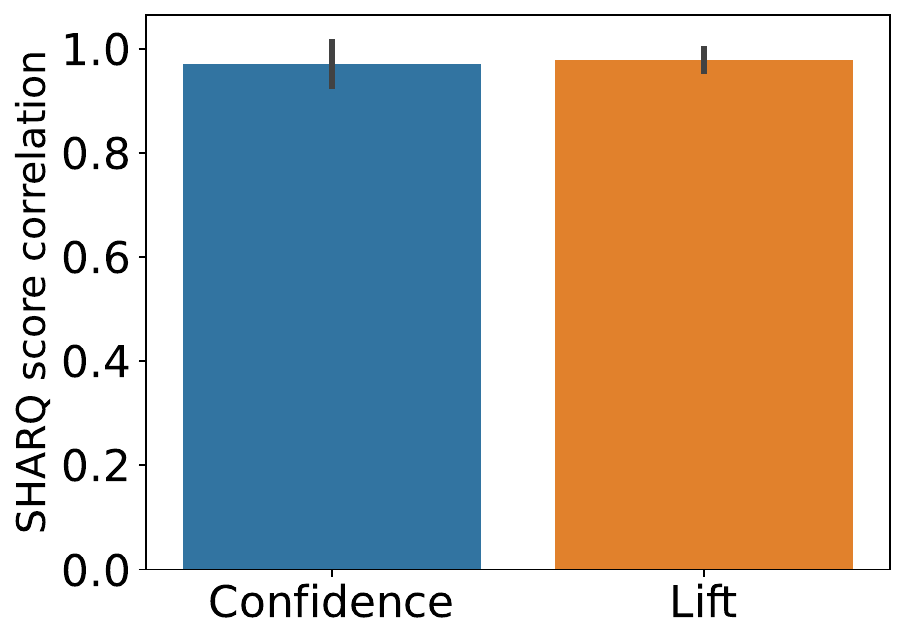}%
        \caption{\rev{Alternative interest. notion}}
        \label{sfig:alt_int}
    \end{subfigure}%
    \hfill
    \begin{subfigure}[b]{0.48\columnwidth} 
        \centering
        \includegraphics[width=\linewidth]{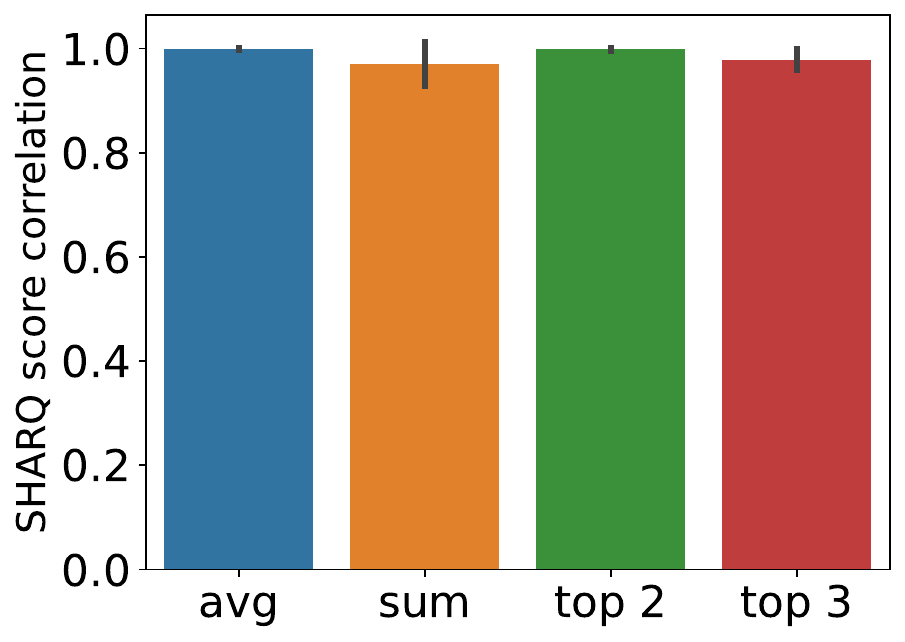}
        \caption{\rev{Alternative agg. interest.}}
        \label{sfig:alt_agg}
    \end{subfigure}%
    \caption{\rev{\system{} Scores correlation when using alternative interestingness notions.}}
    \label{fig:ranks_corr_exsp}
\end{figure}

\subsection{\system{} with alternative notions for rules interestingness}
\label{ssec:ablation}
Finally, we consider the effect of using alternative definitions of rules interestingness on the \system{} scores. 
Recall that our default configuration uses the IS measure~\cite{tan2000interestingness} and $I(R) \coloneqq max_{r \in R}~\text{score}(r)$ for the aggregated interestingness of a rules set, as defined in Section~\ref{ssec:defs}.

To gauge this effect, we examine the correlation between element rankings generated by the \system{} score using different interestingness notions for rules and rule sets. For single rule interestingness, we used (a) \textit{confidence} and (b) \textit{lift}~\cite{brin1997beyond} as alternative functions. For rule sets interestingness, we considered (i) \textit{sum}, in which we sum the scores of the rules in the rules set $R$, $\Sigma_{r\in R} score(r)$; (ii) \textit{Top-2} and (iii) \textit{Top-3} which return the sum of scores of the top two and three most interesting rules, respectively;  and (iv) \textit{average}, calculated as the mean score across the rules in $\rulesdf$.

Figure~\ref{fig:ranks_corr_exsp} shows the rank correlation scores, when using alternative interestingness and aggregated interestingness notion, compared to our default configuration. The rank correlation scores are averaged across all \rev{67} instances in our evaluation sets (.95 confidence intervals are depicted in the figure).

Figure~\ref{sfig:alt_int} shows the rank correlation scores when using alternative interestingness measures. Observe that both confidence and lift scores demonstrate a high rank correlation (0.97 and \rev{0.978}, resp.). 

Figure~\ref{sfig:alt_agg} shows the correlation scores for each alternative aggregated interestingness scores. The correlation scores range from a low of \rev{0.97} for the \textit{sum} function to a high of 0.99 for the \textit{average}. 

This shows that the \system{} scores are robust and consistent across common definitions for rules and rule set interestingness.

%%%%%%%%%OLD:
%In this series of experiments, we investigated the correlations between element rankings generated by the \system{} score using various score functions and interestingness functions. Figure 4 illustrates the differences in correlation scores between the default score and interestingness functions we used and their alternatives (The alternatives are detailed in section 6.1).

%Figure 4a presents the correlation scores for each alternative score function compared to the $I_{TOP}$, our chosen default I function. The correlation scores range from a low of 0.95 for the Sum I function to a high of 0.99 for the Average I function. This suggests a strong level of agreement between the default $I_{TOP}$ and the alternative I functions.

%Additionally, Figure 4b shows the correlation scores between the IS score and two other measures: Lift and Confidence. Lift achieved a correlation score of 0.987, while Confidence scored slightly higher at 0.989.

%In general, all alternative interestingness and score functions maintained a high correlation score, consistently exceeding 0.95.

%% file: related.tex
\section{Related Work}
\label{sec:related}

While \system{} is, to the best of our knowledge, the first framework for explaining association rules mined from relational data, there are several lines of related work that are relevant. We begin by describing existing techniques for analyzing a set of mined association rules, showing that explainability for rules is overlooked. We then present previous work on explaining the results of ML models and database queries, clarifying why such solutions do not help with our problem.

\paragraph*{Analytical tools for analyzing and visualizing association rules.}
A plethora of previous work recognizes the challenge of analyzing the results of rule mining tools, which often return thousands of different association rules. To address this, two prominent methods are suggested: (1) \textit{ranking solutions}, which devise dedicated functions to sort the resulting rules by different interestingness criteria; and (2) \textit{visualization solutions}, which provide graphical means to examine and browse through the resulting rules. None of these solutions, to the best of our knowledge, address the explainability problem of assessing the \textit{contribution} of individual elements to the resulting rules.

(1) \textit{Interestingness functions for ranking and pruning association rules.} Numerous techniques have been devised for \textit{ranking} mined rules according to heuristic notions of \textit{interestingness}~\cite{bayardo1999mining,brin1997dynamic,zhang2009survey,freitas1998objective}. These notions provide a numeric score for each mined rule based on factors such as accuracy, conciseness, reliability, peculiarity, and more (see ~\cite{geng2006interestingness,zhang2009survey} for surveys). Our framework utilizes such notions of rule interestingness, but rather than measuring the rules themselves, it explains how each individual element \textit{contributes} to these scores.

(2) \textit{Visualization interfaces for rules.} While one can use the notions of interestingness to prune uninteresting rules, the number of remaining rules may still be too large for manual analysis. To address this, works such as~\cite{huebner2009diversity} suggest keeping a \textit{diverse} set of rules, covering different parts of the dataset; ~\cite{lent1997clustering,jaroszewicz2002pruning} suggest methods for grouping similar rules together, providing a shorter list of more general patterns; and ~\cite{wong1999visualizing,hahsler2017visualizing} describe visual interfaces for analyzing the mined rules in a matrix-like display, allowing users to examine how items (elements) are connected across multiple rules.

Our explainability framework provides a different way of analyzing the rules, showing which \textit{elements} make the rules interesting. This is achieved by measuring the {contribution} of individual elements to the interestingness scores of rules. Such information can be used, as discussed in Section~\ref{sec:applications}, to detect redundant rules (i.e. those containing non-influential elements), as well as to examine the importance of \textit{attributes} to the overall interestingness of the mined rules.

\paragraph*{Explainability solutions for machine learning models} 
As machine learning models become increasingly complex (e.g., ensemble models, neural networks), there is a growing need for \textit{explaining} model decisions~\cite{doshi2017towards}. A multitude of work suggest solutions for post hoc analysis of model predictions~\cite{shrikumar2017learning,sundararajan2017axiomatic,ribeiro2016should,NIPS2017_7062,ribeiro2018anchors} (see \cite{linardatos2021explainable} for a survey). The most prominent method for explaining ML model predictions is by providing an assessment of \textit{feature importance}~\cite{saarela2021comparison,ribeiro2016should,NIPS2017_7062}, by calculating the contribution of each feature to the predictive performance of the model—either locally~\cite{ribeiro2016should,shap_ml} for a single prediction or globally~\cite{zien2009feature,ibrahim2019global} for all model predictions.

In the context of explaining association rules, we follow~\cite{shap_ml}, a highly popular ML explainability framework, and calculate element contribution based on the notion of Shapley Values~\cite{shapley1953value}. However, using Shapley values in our context requires a novel adaptation of the concept as well as dedicated algorithms to speed up the computation, as detailed in Section~\ref{sec: alg}. Without our optimized algorithms, as shown in our experimental evaluation, the calculation of \system{} is infeasible in most rule-mining settings.

\paragraph*{Explainability solutions for database queries}
Explainability has also been studied within the data management community, focusing primarily on explaining the results of database queries~\cite{GKT-pods07,why,chapman2009not,Amsterdamer2011provenance}. This is often done by utilizing data provenance and causality-based notions such as intervention and influence to identify tuples whose existence or absence affects the result of the inspected query.

In particular, \cite{livshits2019shapley,davidson2022shapgraph} also use the notion of Shapley values to calculate the importance of database \textit{tuples} to a given query result set, and devise a dedicated optimization framework to facilitate the expensive computation. Since our work focuses on explaining association rules rather than queries, it requires a different Shapley adaptation and computational framework.

\paragraph*{Assistance tools for Data Analysis.}
In a broader sense,
our work is a part of an ongoing research effort whose goal is to facilitate the difficult task of analyzing data. To this end, works such as~\cite{srinivasan2018augmenting,singhdbexplorer,bao2015exploratory,bespinyowong2016exrank} suggest simplified exploration interfaces that allow users to wrangle the data without explicitly writing queries. Other systems automatically produce data visualizations~\cite{luo2018deepeye,wongsuphasawat2016voyager} and provide users with general, actionable insights~\cite{tang2017extracting,huang2019leri} mined from the data.
However, none of this work addresses the explainability problem of mined rules.

%% file: conclusions.tex
\section{Conclusions}
\label{sec: concl}
\system{} is a novel explainability framework for a set of association rules which measures each element's contribution to the interestingness of the set. The metric is based on the notion of Shapley values, and captures the frequency of the element in the dataset as well as the variability in interestingness across rules of different lengths when the rule is excluded. Since calculating the \system{} score of an element can be extremely expensive, we give an efficient \sysopt{} algorithm that is linear in the size of the rule set, and further reduce the cost for a set of elements in a multi-element \sysopt{} algorithm. We also show two additional use-cases, rule importance and attribute importance.
Extensive experiments show the effectiveness of this approach. 

In future work, we plan to extend our approach to decision and classification rules, resulting from predictive ML models. We also plan to look at the problem of updates to the data/rule set, and whether SHARQ scores can be incrementally updated.

%\susan{Other ideas are in an eat environment.}
\eat{
Future work:  
Hammer out user cases better;
other questions in explainability other than feature importance (e.g. counter factuals); additional use case, calculating SHARQ within the context of a subset of rules and comparing against the SHARQ score over entire set of rules; what to do in the case of updates, incremental updates to SHARQ scores
} 
\eat{More ideas: extend to decision/classification rules. I.e., ML/predictive models that are composed or rules. } 

\eat{threshholds on support and interestingess in rule mining algorithms, produce rules that are not as large as |A|; can also set a threshold on rule length to $\tau$; in our case the max rule size is less than 10, even when the number of elements is large (40)}